\documentclass[10pt,journal,compsoc]{IEEEtran}

\usepackage{color}
\usepackage{hyperref}           
\hypersetup{
    colorlinks=true,
    linkcolor=blue,
    filecolor=red,      
    urlcolor=magenta,
    breaklinks=true,            
}
\usepackage{breakurl}           

\usepackage{booktabs} 
\usepackage{listings} 
\usepackage{graphicx}
\usepackage{array}
\usepackage[framemethod=TikZ]{mdframed}
\mdfdefinestyle{MyFrame}{%
    linecolor=black,
    outerlinewidth=1pt,
    roundcorner=5pt,
    innertopmargin=5pt,
    innerbottommargin=5pt,
    innerrightmargin=5pt,
    innerleftmargin=5pt,
    backgroundcolor=gray!10!white}
\usepackage{ragged2e} 
\usepackage{graphicx}
\usepackage{amsmath}
\usepackage{cleveref}
\crefname{section}{Section}{Section}
\Crefname{section}{Section}{Section}
\crefformat{section}{§#1#2#3}

\usepackage[T1]{fontenc}
\usepackage{subfig}
\usepackage[htt]{hyphenat}
\usepackage{pifont}
\usepackage{hyperref}
\usepackage{wrapfig, mwe}
\usepackage{amssymb}
\usepackage{amsmath,wasysym}
\usepackage[utf8x]{inputenc} 
\usepackage[referable,flushleft]{threeparttablex}
 
\usepackage{booktabs}
\usepackage{multicol}
\usepackage{siunitx}
\bstctlcite{IEEEexample:BSTcontrol}

\usepackage{amsthm}
\newtheorem{definition}{Definition}

\usepackage[]{xcolor}
\newcommand{\TODO}[1]{\textcolor{red}{TODO: #1}}\newcommand\todo\TODO
\newcommand{\paul}[1]{\textcolor{black}{#1}}
\newcommand{\paulm}[1]{\textcolor{black}{#1}}

\usepackage{tikz}
\newcommand\encircle[1]{%
\tikz[yshift=-2pt] 
   \node (X) [draw, shape=circle, inner sep=0, scale=0.5, fill=black, text=white] {\strut #1};}
   
\usepackage{pifont}

\usepackage{xspace}
\newcommand{\sysname}{{\small\textsc{IntRepair}}\xspace}
\usepackage{lastpage}
\usepackage{fancyhdr}
\bstctlcite{IEEEexample:BSTcontrol}

\ifCLASSOPTIONcompsoc
  \usepackage[nocompress]{cite}
\else
  \usepackage{cite}
\fi
\ifCLASSINFOpdf
\else
\fi
\begin{document}
\sloppy                         
\title{\textsc{IntRepair}: Informed Repairing of Integer Overflows}


\author{Paul Muntean\href{https://orcid.org/0000-0002-2462-7612}{\includegraphics[scale=0.7]{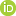}},
Martin Monperrus\href{https://orcid.org/0000-0003-3505-3383}{\includegraphics[scale=.7]{figs/orcid.png}}, 
Hao Sun\href{https://orcid.org/0000-0001-8634-2367}{\includegraphics[scale=.7]{figs/orcid.png}}, 
Jens Grossklags\href{https://orcid.org/0000-0003-1093-1282}{\includegraphics[scale=.7]{figs/orcid.png}}, and 
Claudia Eckert

\IEEEcompsocitemizethanks{
\IEEEcompsocthanksitem P. Muntean was with the Department of Computer Science, Technical University of Munich, Germany. E-mail: paul.muntean@sec.in.tum.de

\IEEEcompsocthanksitem M. Monperrus is with the Department of Computer Science, KTH Royal Institute of Technology, Sweden. 
E-mail: martin.monperrus@csc.kth.se

\IEEEcompsocthanksitem H. Sun was with the State Key Laboratory of Novel Software Technology, Department of Computer Science and Technology, Nanjing University, China. E-mail: shqking@gmail.com

\IEEEcompsocthanksitem J. Grossklags is with the Department of Computer Science, Technical University of Munich, Germany. E-mail: jens.grossklags@in.tum.de

\IEEEcompsocthanksitem C. Eckert is with the Department of Computer Science, Technical University of Munich, Germany. E-mail: claudia.eckert@sec.in.tum.de

}

}

\IEEEtitleabstractindextext{%
\justify
\begin{abstract}
Integer overflows have threatened software applications for decades.
Thus, in this paper, we propose a novel technique to provide automatic repairs of integer 
overflows in \texttt{C} source code.
Our technique, based on static symbolic execution, fuses 
\textit{detection}, \textit{repair generation} and \textit{validation}.
This technique is implemented in a prototype named \sysname. 
We applied \sysname to 2,052 \texttt{C} programs (approx. 1 million lines of code) contained in SAMATE's Juliet test suite and 50 
synthesized programs that range up to 20 KLOC.
Our experimental results show that \sysname is able to effectively detect integer overflows and successfully repair them, 
while only increasing the source code (LOC) and binary (Kb) size by around 1\%, respectively. Further, we present the 
results of a user study with 30 participants which shows that \sysname repairs are more than 10x
efficient as compared to manually generated code repairs.

\end{abstract}

} 

\maketitle

\IEEEdisplaynontitleabstractindextext

\IEEEraisesectionheading{\section{Introduction}\label{intro}}

Integer overflows are a well-known cause for memory corruptions (\textit{e.g.,} \texttt{C++} object type confusion \cite{castsan}) which could lead to buffer overflows (IO2BO) and potentially to control-flow hijacking attacks \cite{tauCFI}. Integer overflows are a widely known type of vulnerability~\cite{intscope} that has threatened programs for decades.
\paul{It now even has a revival, with the detection of integer overflows in Ethereum's Solidity smart contracts \cite{osiris}}. To address this challenge, it has been, for example, proposed to apply the concept of automatic repair \cite{Monperrus2015} (see our Definition \ref{defx}) to fix integer overflows. This idea consists of employing a system that produces code repairs in order to fix integer overflows. The most recent systems are dynamic: for instance, 
SoupInt~\cite{soupint} is a runtime-based binary repair generator. On the contrary, we aim at having a fully static approach, with no need for defining or generating inputs that trigger the integer overflow.

In this paper, we tackle these two problems in a novel system, \sysname\footnote{Source code:~\url{https://github.com/TeamVault/IntRepair}}, usable for automatic repair of integer overflows.
Our key idea is to use symbolic execution~\cite{klee, intscope, cadar2013symbolic, avgerinos2014enhancing, marinescu2012make}  to reason about the repair.
\paul{More specifically, our key contribution compared to previous work is that fault detection\footnote{Fault detection demo:~\url{https://goo.gl/uNvdRp}} works hand in hand with program repair\footnote{Fault repair demo:~\url{https://goo.gl/912Jux}}.}
In particular, we fuse the fault localization and repair generation phases into a single algorithm via SMT solving. We present a novel integer overflow repair technique that realizes this idea and the corresponding prototype tool, \sysname. To the best of our knowledge, \sysname is the first approach for automated repair of integer overflows that does not require test cases.

Given a faulty program, \sysname generates an SMT system that captures integer overflow manifestation and allows for valid repair (see our Definition \ref{def1}) and correct repair (see our Definition \ref{def2}) generation. This is achieved by interweaving the fault detection 
SMT constraints with newly synthesized constraints used for correct fault generation.
\paulm{
Further, the integer overflow detection component on which \sysname relies does not generate false negatives (\textit{i.e.,} if there is a genuine integer overflow then this will be detected). Further, we do not distinguish between intended or unintended integer overflows, yet false positives may happen due to certain implementation limitations, such as loop unrolling or recursion depth. Our goal is to catch all faults. Further, as far as we are aware of, there is no tool currently capable to distinguish between intended and unintended integer overflow faults. Note that for this reason we leave the final repair insertion decision to the programmer as only he knows the exact intended program behavior. Lastly, in this way he may opt to insert or not the (otherwise) fully automated synthesized repair.}

\sysname's implementation is based upon the Codan static symbolic execution engine \cite{safecomp, ibing:hase, ibing:date}, which uses the Z3 SMT solver \cite{z3} to solve the extracted constraints.
Note that our repair generation technique is general enough to be implemented on top of other symbolic execution engines as well.
We evaluated \sysname on 2,052 \verb!C! programs contained in the SAMATE's \verb!C/C++! benchmark suite~\cite{juliet},
totaling more than one million lines of code (LOC). The evaluated programs contain all possible program control flows that may lead to an integer overflow. 
Further, we use a synthesized benchmark containing 50 large \verb!C! programs (up to 20 KLOC) seeded with integer overflows.
Our experimental results show that \sysname is able to:
(1) effectively detect all integer overflows, and
(2) successfully repair the programs at the source code level. 
In addition, we present the results of a user study with 30 participants showing 
that \sysname is more effective compared to traditional debugging based program repair.
To the best of our knowledge, there are no other open source tools with which we can compare against. In particular, DirectFix \cite{directfix}, Angelix \cite{angelix}, or SemFix \cite{semfix} are not suitable for comparison purposes as these tools use test cases to locate the fault: on the contrary, we do not assume the presence of test cases. 

In summary, we make the following contributions:

\begin{itemize}
\item \textbf{Repair Technique.} We design a novel source code repair generation technique for integer overflows in \verb!C! programs.

\item \textbf{Repair Tool.} 
We implement \sysname, as a prototype of our novel integer overflow repairing technique, for \verb!C! source
code programs. It can be automatically used to repair integer
overflows for multiple integer precisions.

\item \textbf{Quantitative Evaluation.} We evaluate \sysname thoroughly with 2,052 \verb!C! programs contained in SAMATE’s
Juliet standard test suite for \verb!C/C++! source code. We also evaluate it with 50 synthesized programs which range up to 20 KLOC. We show that \sysname's code repairs are effective and induce low runtime overhead. 

\item \textbf{Controlled Experiment.} We evaluate \sysname within
a controlled experiment with 30 participants and show that
\sysname is more time-effective than manually repairing the same
programs.

\end{itemize}

\textbf{Outline.} The paper is organized as follows. 
In~\autoref{sec:Background}, we present the required background knowledge needed in order to understand the rest of this paper. In~\autoref{Overview}, we present the overview of our technique. In~\autoref{sec:design}, we highlight design and implementation details of \sysname.
In~\autoref{sec:evaluation}, we evaluate \sysname, and
in~\autoref{Discussion}, we discuss limitations of \sysname. In~\autoref{sec:relatedwork}, we review related work, and in~\autoref{sec:conclusion}, we offer concluding remarks.

\section{Background}
\label{sec:Background}
In this section, we present background information.

\subsection{Integer Overflows}
\label{Integer Overflow Detection Techniques}
Integer overflow is a known cause of memory corruption and a widely known type of vulnerability as mentioned by Wang \textit{et al.}~\cite{intscope}. It often leads to stack or heap overflow and thus
is usually exploited indirectly, as opposed to buffer overflows which are exploited directly, see Brumley \textit{et al.} \cite{rich}. More specifically, integer overflows occur at runtime, when the result of an integer expression exceeds the maximum allowed value (\textit{i.e.,} $2^{32}-1$).

\begin{figure}[ht]
    \centering
    \includegraphics[width=70mm]{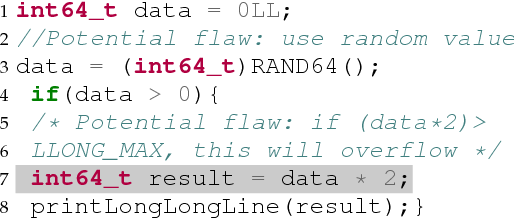}
    \captionof{figure}{Integer overflow shaded gray at line seven.}
    \label{Integer overflow based memory corruption example.}
\end{figure}
\autoref{Integer overflow based memory corruption example.} depicts an integer overflow memory corruption at line number 7, which could manifest 
because there is no proper check in place for verifying the range of admissible values for \texttt{data}. 
This integer overflow (and potential underflow) fault can be avoided by checking the value of \texttt{data} to see if it is 
less than or equal
to $\frac{LLONG\_MAX\_VAL}{2}=4.611.686.018.427.387.903 \approx 4^{18}$ and
greater than or equal
to $\frac{-LLONG\_MAX\_VAL}{2}$.

\subsubsection{Characteristics of Integer Overflows} 
Integer overflows can be classified as malicious or benign. 
Essentially, an integer overflow manifests itself, when the program receives user-supplied input and subsequently the input value is used in an arithmetic operation to trigger an integer overflow.
Thus, a smaller than expected value is supplied to the memory 
allocation function and as a result a smaller than expected memory will be allocated. 
Deciding between the types of integer overflow related problems is rather difficult and 
a lot of research has been devoted in the last years to this type of classification \cite{inteq}. The general desire in the research community is to categorize different hard to find integer overflows w.r.t. how exploit-prone these are in contrast to just finding and repairing them.

\subsubsection{Problems Due to Integer Overflows}
\label{Integer Overflow Related Problems.} 
In this section, we review the integer overflow related problems that we address within \sysname and briefly describe other related types. \paul{More precisely, in this paper, we address only integer overflow, as described in document CWE-190~\cite{cwe190}, and integer underflow~\cite{cwe191:underflow}}, which
is the result of multiplying two values with each other and the result is less than the minimum admissible integer value. This is due to the fact that the product subtracts one value from another.

Further, other integer related problems are:
CWE-192, integer coercion error~\cite{cwe192:coercion},
manifests during bad type casting and extension or truncation of primitive data types.
CWE-193, off-by-one error~\cite{cwe193:offbyone}, manifests
during product calculation/usage; an incorrect maximum/minimum value is used which is one more, or one less, than the correct value.
CWE-194, unexpected sign extension~\cite{cwe194:sign}, appears when
an operation performed on a number can cause it to be sign-extended when it is transformed into a larger data type.
CWE-195, signed to unsigned conversion error~\cite{cwe195:signed}, manifests when
a signed primitive that is used inside a cast to an unsigned primitive can produce an unexpected result if the value of the signed primitive cannot be represented using an unsigned primitive.
CWE-196, unsigned to signed conversion error~\cite{cwe196:unsigned}, manifests when
an unsigned is used inside a cast to a signed primitive, which can produce an unexpected value if the result of the unsigned primitive cannot be represented using a signed primitive.
CWE-197, numeric truncation error~\cite{cwe197:truncation},
manifests when a primitive is cast to a primitive of a smaller size and data is lost in the conversion.
CWE-680, integer overflow to buffer overflow~\cite{cwe680},
appears when an integer overflow occurs that causes less memory to be allocated than expected, which can lead to a buffer overflow.

\paulm{
\subsection{Repairing Integer Overflows}
\label{How to Repair an Integer Overflow Memory Corruption?}
There are several approaches to repair integer overflow based memory corruptions. Most of these techniques are based on input validation.
Of all these techniques, manual fixing is the most tedious and provides fewer guarantees than other techniques. Compiler-based approaches add repairs in all fault prone locations and scale well to large code bases. As such, we next present symbolic execution based input validation and compare it against other techniques.
}
\paulm{
As compared to other approaches, symbolic execution based techniques can be used to systematically check program paths for integer faults and to propose repairs. As a consequence, the repairs are cheap to construct and to insert. On the one hand, symbolic execution-based techniques can achieve more guarantees than other repair generation techniques. On the other hand, these techniques are based on computationally intensive analysis strategies which, if not applied in a suitable manner, may not scale well (or at all) with large programs. For this reason, we believe that repair tools should be used  by programmers early during development since the level of software complexity is low and it increases as the number of code lines becomes larger. Lastly, we believe that in general:
(1) manually written source code repairs, the industry standard, should be avoided and only used in \textit{easy} to address situations \cite{search:repair},
(2) compilers should not be used for repairing integer overflows since the number of guarantees which these can offer is low, and
(3) specialized tools which provide more guarantees should be used for repair generation.}

\section{Overview}
\label{Overview}
In this section, we present the overview of our approach. 

\subsection{Repair Definitions}
\label{Repair and Validation Defnitions}
In this section, we provide the definitions of an automatic software repair as presented by Monperrus \cite{Monperrus2015}, and then of \sysname's valid repair and correct repair.

\begin{definition}[Software Repair]
\label{defx}
Automatic software repair consists of automatically finding a repair $r$ to a software fault identified by an oracle $o$, without human intervention, and which applied modifies one or more lines of source code.
\end{definition}

Our work is based on a workflow with two characterizations of repair: valid repairs (initial) and correct repairs (stronger) which we now define.

\paulm{
\begin{definition}[Valid Repair]
\label{def1}
Let $r$ be a software repair for program $P$ and a fault $f$, $v$ is a validation oracle of $r$, $I$ is the set of inputs on which $P$ behaves as intended and which does not generate $f$, and $B$ is the set of valid behaviors of $P$. Then $r$ is said valid under $v$ iff:
\begin{itemize}
\item $r$ enforces $v$
\item $r$ does not change $B$ for $I$.
\item $r$ changes $B$ for non-standard $I$.
\end{itemize}
\end{definition}
}

\paulm{
\begin{definition}[Correct Repair]
\label{def2}
Based on the same terminology, let $c$ be a correctness oracle for software repair $r$ of program $P$ for a fault $f$. Let $Q$ be the set of all reachable program paths of $P$ and a program path $p$ $\in$ $Q$. Then $r$ is correct under $c$ iff:
\begin{itemize}
\item $r$ enforces $c$
\item $r$ removes $f$ on all $p$ $\in$ $Q$ that reach $f$.
\item $r$ does not introduce other behavior than $B$, \textit{i.e.,} $r$ removes $f$ without introducing new faults.
\end{itemize}
\end{definition}
}

\paulm{These definitions underlay \sysname's automatically generated repairs w.r.t. to their fundamental properties. In this paper, we use symbolic execution to identify valid and correct repairs.}

\subsection{Symbolic Execution Engine}

\paulm{In this work, we use a symbolic execution based approach} for fault detection and repair generation. A symbolic execution engine~\cite{safecomp, issa} constructs a control flow graph (CFG) for each analyzed program and extracts execution paths. 
Constraints along the execution path are encoded into SMT equations, using the SMT-LIB format~\cite{smtlib}.
The translation of CFG nodes into SMT is performed by a translator algorithm, which extends the program's abstract syntax tree (AST) visitor class, according to the visitor pattern~\cite{gamma:patterns}. 
This is usually based on a \textit{bottom-up} traversal of each 
program statement located on the currently analyzed program execution path. 

Further, we use the Codan static symbolic execution engine~\cite{safecomp, ibing:hase, ibing:date}.
In Codan, single static assignment (SSA) variables are created for \verb!C! expressions, which are associated with no variables in the analyzed program.
Before creating a new variable, the interpreter checks whether there is already a symbolic variable.
Further, for all SMT formulas created for a particular program statement, one symbolic variable is created for each of the variables contained in the original program statement. Next, a single path is extracted from the previously computed CFG and traversed. In Codan: (1) loops can be traversed a configurable number of times, (2) the analysis can be customized to look, for example, for the first \textit{N} faults located on the currently analyzed program path, and (3) Codan performs path-caching and backtracking traversal, which avoids traversing the whole program path from the beginning and collecting all constraints again.
Codan uses the Z3 \cite{z3} solver as its backend in order to solve SMT constraints.

\subsection{Main Fault Localization Engine Features}

\textit{\textbf{Unrestricted Context Depth.}} The symbolic execution engine performs an inter-procedural path sensitive analysis with a call string approach. For each function a CFG is built.

\medskip\noindent\textit{\textbf{Loop Unrolling Trade-offs.}} Each program loop can be unrolled up to a certain depth by setting a specific threshold value. Currently, we unroll each loop up to 10 times. Note that we experimented with unrolling each loop also 100 and 1000 times and for the tested programs this made no difference. Further, this value is configurable by the user.

\medskip\noindent\textit{\textbf{Library Calls.}} Our approach can handle library calls (\textit{e.g.,} \texttt{memset}, \texttt{memcpy}, etc.) by providing upfront for each of these functions a stub SMT function, which models the various function behaviors.

\medskip\noindent\textit{\textbf{Finding Program Paths.}} We use a fixed deterministic thread scheduling algorithm for running the symbolic execution. The symbolic execution is run with approximate path coverage which uses Depth-First Search (DFS). During DFS, program states are backtracked and branch decisions are changed.

\medskip\noindent\textit{\textbf{Automatic Slicing.}} We perform automatic slicing over the data flow in order to verify conditions on a program path and over the control flow to separate the analysis of different program paths. 

\medskip\noindent\textit{\textbf{Context Sharing for Different Checkers.}} Our engine allows that multiple checkers (\textit{e.g.,} integer overflow, buffer overflow, race condition, etc.) run in parallel during symbolic execution. Checkers are allowed to share the contexts because there is separation from the symbolic path interpretation.

\medskip\noindent\textit{\textbf{Logic Representation.}} We use the SMT-LIB sub-logic of arrays, uninterpreted functions and non-linear integer and real arithmetic (AUFNIRA) since this approach can be automatically decided. Pointers, for example, are handled as symbolic pointers by the engine interpreter with a target and a symbolic integer as offset formula and output as logical formulas when dereferenced. 

\medskip\noindent\textit{\textbf{Path Validation.}} The program execution path validation is triggered at branch nodes and uses the same interface as the checkers. For all path decisions up to the current branch, the path validator queries the equation SMT-LIB linear system slice. The path validator throws a path unsatisfiable exception if the solver answer is unsatisfiable; then the symbolic execution proceeds with the next path. 

\medskip\noindent\textit{\textbf{Satisfiability Modulo Theories Solving.}} IntRepair uses SMT solving for three main purposes. First, for checking satisfiable and not satisfiable program execution paths. Second, for checking the presence of an integer overflow, additional SMT constraints are added to the SMT linear constraint system which was used to check path satisfiability in order to check for the presence of an integer overflow. Third, for checking if a repair removes an integer overflow, additional SMT constraints are added to the SMT linear constraint system which was used to check path satisfiability.

\medskip\noindent\textit{\textbf{Path-sensitive Tracing of Shared Variables.}} It is possible to use shared variables between threads in a context-sensitive way. This is accomplished by first marking all global shared variables, and then the shared property is inferred over data flow constructs such as references, assignments, function call parameters and return values. 

\medskip\noindent\textit{\textbf{Deep Nested \texttt{C} Structs.}} The \texttt{C} language program statement to SMT translator engine component is designed to recursively traverse deep nested \texttt{C/C++} structures in order to determine the real type of a field inside the struct. Accordingly, our tool does not lose precision when dealing with \texttt{C} struct data types.

\subsection{Overflow and Underflow Checks}
\label{Overflow and Underflow Checks}
In the following, we present the three types of overflow and underflow checks that are supported by \sysname. If one of the preconditions evaluates to true, then an integer overflow has been found. \paul{We selected these three operations since these solve a large proportion of integer overflows \cite{ioc, ioc:icse}. We plan to address the detection of integer overflows with more sophisticated detection schemes in next iterations of our tool. Next, we present the checks used by \sysname, which are modeled as the following  preconditions.}

\medskip\noindent\textbf{\textit{First precondition.}} The addition of two integers, in which one is a variable and the other is a positive constant, will lead to an integer overflow
if the following expression evaluates to true:
  $((s_{1} > 0) \land (s_{1} > (\text{INT\_MAX} − s_{2})))$.
  Note that $s_{2}$ is the positive constant.

\medskip\noindent\textbf{\textit{Second precondition.}} The multiplication of two integers, in which one is a variable and the other is negative, will lead to an integer overflow or
underflow if the following expression evaluates to true:
    $((s_{1} > 0) \land (s_{1} > (\text{INT\_MIN} / s_{2}))) \lor ((s_{1} < 0) \land (s_{1} < (\text{INT\_MAX} / s_{2})))$.
 Note that $s_{2}$ is the negative constant.

\medskip\noindent\textbf{\textit{Third precondition.}} The multiplication of two equal integers will lead to an integer overflow 
  or underflow if the following expression evaluates to true:
   $((s_{1} > 0) \land (s_{1} > (\text{sqrt(}\text{INT\_MAX}\text{)})))$ $\lor$  
   $((s_{1} < 0)$ $\land$ $(s_{1} < (-\text{sqrt(}\text{INT\_MAX}$ $\text{)})))$.

%
%
%
%
The above preconditions are used over multiple types of inputs for $s_{1}$ or $s_{2}$ (\textit{e.g.,} \texttt{RAND32()}, this is the wrapper for the \verb!C! random function, \texttt{fscanf()} etc.).
The preconditions can be applied over multiple integer precisions, meaning that \text{INT\_MAX} can take different values depending on the currently used integer precision 
in the analyzed program.
We call this value the maximum admissible upper bound value of an integer. This value is determined automatically during program analysis.
Further, the variables $s_{1}$ or $s_{2}$ can take different types: \texttt{char}, \texttt{int64\_t}, \texttt{int}, \texttt{short}, \texttt{unsigned int}. 
In contrast to IOC's preconditions, which aim to avoid an unconfirmed integer overflow during program runtime, 
our preconditions help to guide repairs at the right code location, where the integer 
overflow was detected and confirmed. This results in avoiding the fault during runtime. Note that we always use symbolic execution again as a last step to confirm 
that the fault was completely fixed after a repair was inserted.

Other operations, which may lead to integer overflows, are: truncation, bit shifts and subtraction operations (see Figure 6 in Pereira \textit{et al.} \cite{io:conditions} for more details).
Currently, \sysname does not support these operations, but we plan to support them in an updated version of \sysname.

\begin{figure*}[ht!]
      \centering
      \includegraphics[scale = 0.7, resolution = 100000, width=2.0\columnwidth]{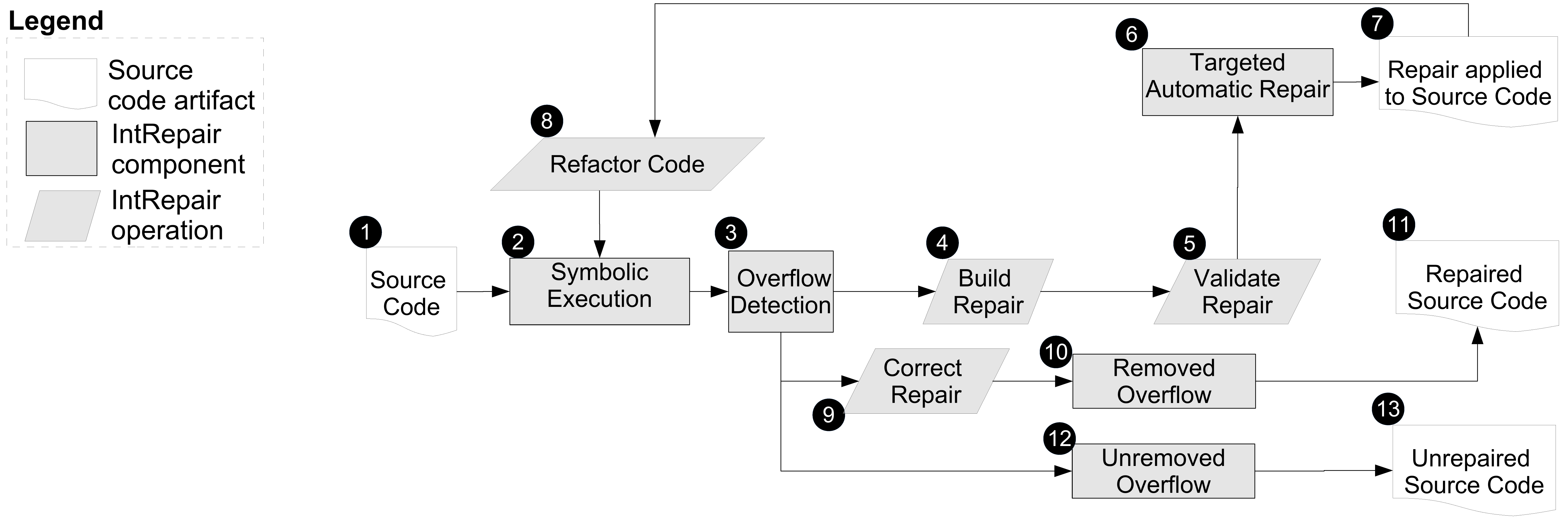}
      \caption{\paul{Repair generation overview.
      The developer provides the faulty program as source code (1) which is next analyzed with the help of the symbolic execution engine (2) and depending on the result of integer overflow detection (3), next
      a repair is constructed (4) which represents a valid repair (5), next the steps (6) - (8) are followed in order to apply the generated repair and refactor the source code after applying the repair. Lastly, depending on step (3), either steps: correct repair (9),
      removed overflow confirmation (10), and repaired source code (11), or step (12) unremoved overflow, and step (13) unrepaired source code are executed.}}
      \label{Overview of IntRepair}
\end{figure*}

\section{Design and Implementation}
\label{sec:design}

\subsection{Workflow of \sysname}
\label{sec:overview}

\autoref{Overview of IntRepair} 
provides an overview of {\sysname} showing its main components.
Initially, the 
{\tiny\encircle{\Large{1}}} \textit{source code} is passed into the 
{\tiny\encircle{\Large{2}}} \textit{symbolic execution} engine, which first constructs a CFG
and \paul{then extracts program execution paths from this CFG.\footnote{A certain maximum recursion depth for each analyzed path is specified as configuration parameter.}}
Program execution constraints are collected along each 
path 
{\tiny\encircle{\Large{3}}} and the \textit{integer overflow detection} component 
identifies whether an integer overflow might occur for one of the given execution paths. 
In {\tiny\encircle{\Large{4}}} a repair pattern is selected in order to build a repair.
The {\tiny\encircle{\Large{5}}} validate repair (see Definition \ref{def1} for valid repair) step is performed. During this step
it is checked whether the negated
SMT constraints invalidate the previously detected integer overflow fault.
If this is the case, in
{\tiny\encircle{\Large{6}}} a {\textit{targeted automatic repair}} is generated that potentially removes the previously detected fault. 
After applying the repair, we obtain the 
{\tiny\encircle{\Large{7}}} {\textit{repaired source code}}, and next in 
{\tiny\encircle{\Large{8}}} the {\textit{refactored code}} which is again validated through {\tiny\encircle{\Large{2}}}. Further, 
in case {\tiny\encircle{\Large{3}}} {\textit{overflow detection}} does not find an overflow, \sysname produces
a correct repair {\tiny\encircle{\Large{9}}}, and it confirms the
{\tiny\encircle{\Large{10}}} {\textit{removed overflow}}, indicating that the fault was successfully 
removed and
{\tiny\encircle{\Large{11}}} {\textit{repaired source code}} has been synthesized.
Further, in case the integer overflow was not removed, then \sysname produces 
{\tiny\encircle{\Large{12}}} an {\textit{unremoved overflow}} report, 
and the result is {\tiny\encircle{\Large{13}}} \textit{unrepaired source code}. 
\paulm{Further, note that we only ensure that the previously detected integer overflow was repaired. The program may still be exposed to other types of faults which were potentially hidden/masked by the fault which was just repaired.}

Lastly, note that for verifying a repair after it was applied, \sysname does not require any test case since the integer overflow component can decide if the fault was removed. However, a test case \textit{can} be used in conjunction with the fault detection component in order to confirm that the fault was removed from the source code location where it was first detected before repairing.


\subsection{Overflow Detection}
\label{Repair Location Search}

In order to generate code repairs, \sysname initially needs to detect the precise location where the integer overflow resides in the program. This is the goal of \sysname's repair location search algorithm, which we now present. First, \sysname constructs the CFG of the analyzed program. Next, the following steps are performed consecutively in order to detect a fault.

 (1) Each program execution path is extracted from the previously generated program CFG.
 
 (2) The extracted path is traversed and path satisfiability checks are performed at branch nodes.

 (3) When \sysname encounters an integer fault prone code location (\textit{e.g.,} assignment statement) on the analyzed 
path, an integer overflow check is performed by notifying the interpreter.
 
 (4) The notification is delegated to the appropriate checker (\textit{e.g.,} integer overflow checker) by the interpreter.
 
  (5) The slice of SMT equations of the symbolic variable, which potentially may overflow, together with  
corresponding integer overflow satisfiability checks is queried by the integer overflow checker.
 
  (6) The check verifies if the symbolic variable, which potentially caused the integer overflow, can be greater (\paul{if this is true, then an integer overflow is reported by providing its fault identification information which contains the key information, in particular the  line number of the fault)} than the currently used integer upper bound value (\textit{i.e.,} \texttt{INT\_MAX}). These upper bound values are extracted from the \verb!C! standard library contained in the \texttt{limits.h} file.
The lower bound is obtained by negating the currently used upper bound value and and adding 1 to the result.

  (7) In case the SMT solver replies \texttt{SAT} (satisfiable, integer overflow fault present) to the previously submitted SMT query, then a problem report 
with problem ID (unique system string, the ID refers to which checker detected the fault), the file name where the fault 
was detected, and the line number in the file where the fault is located will be created and stored.

\paulm{
\textit{\textbf{Optimization.}}
\sysname searches for integer overflow locations on each analyzed path which are most fault prone to integer overflow such as assignment or multiplication statements and discards for now other places. Further, due to \sysname’s backtracking capabilities, see Section 2.2 paths are not always re-analyzed from the start. Essentially, during path analysis our engine, in case of a not satisfiable path, backtracks up to the next upper branch node. Next \sysname starts to analyze an alternative path reusing part of the SMT constraints which it has gathered up to the path branch node. This avoids recomputing all path constraints for that analyzed path from the beginning of the path. Lastly, this optimization helps to save a lot of runtime overhead (approx. 50\% according to our experience).}

\subsection{Repair Patterns}
\label{sec:patterns}
In this section, we present the repair
patterns available in \sysname. We explain how these patterns can help to repair integer overflows and how the patterns can produce correct repairs.
Repair patterns are stored in a decision tree. This decision tree is, before the fault repair analysis is started (before using \sysname), manually constructed by \sysname's maintainer and stored in a tree data structure for later usage. 
The tree construction algorithm is based on several types of \verb!C! language statements. More specifically, the components of these statements, are decomposed into AST trees and mapped onto the tree nodes. As such, we have from the root node to the leave nodes the operator used in the faulty statement, the type of operations, the type of the result variable, and other information such as if the operands are constant variables and if these, for example, have side-effects.
The main rationale behind the decision tree is to help \sysname perform a time-efficient search for each detected fault and to pick a core repair that fits best.

\begin{figure*}[ht!]
      \centering
      \includegraphics[resolution=100000, width=2.10\columnwidth]{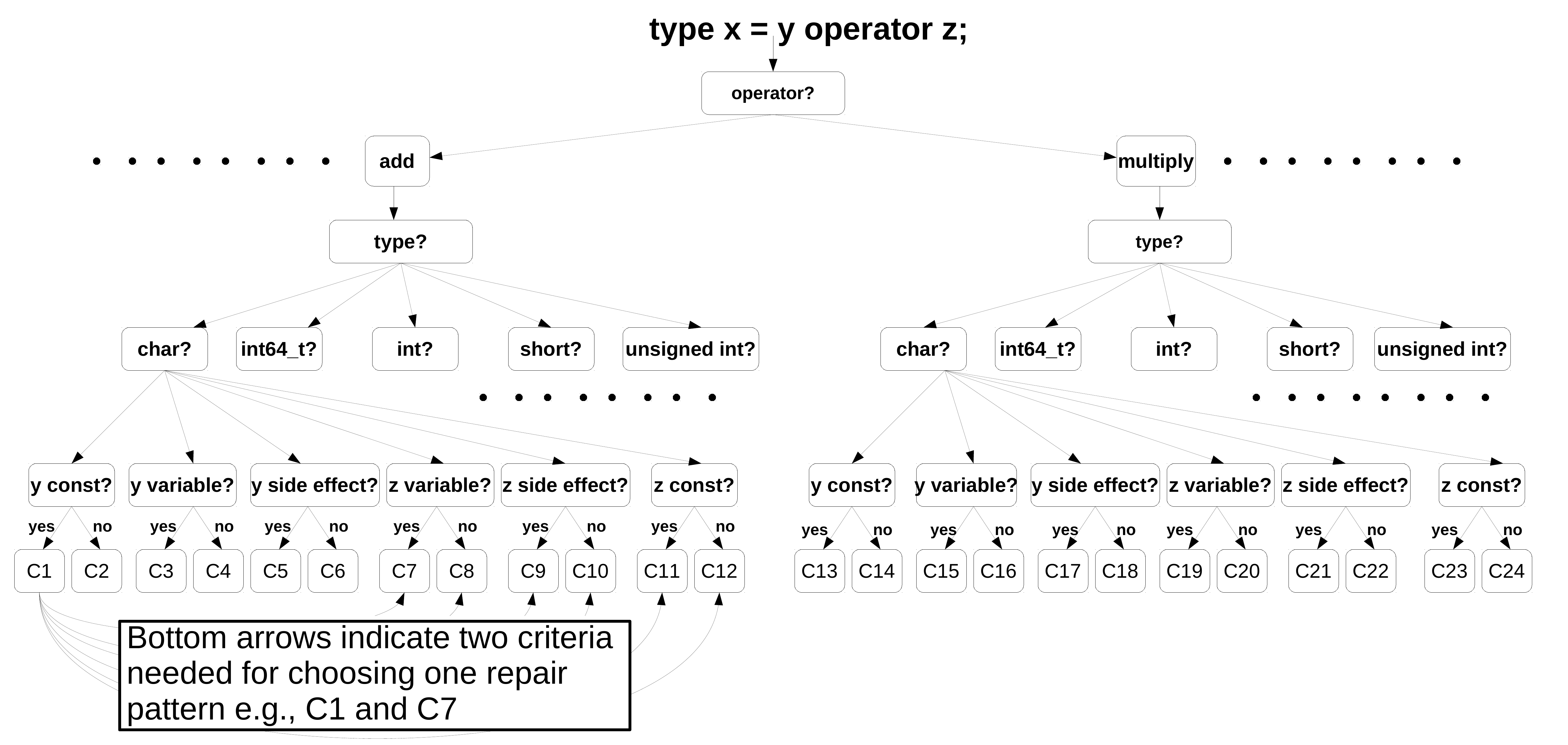}
      \caption{Decision tree used by \sysname for storing and retrieving repair patterns.}
      \label{Overview of repair pattern selection}
\end{figure*}
\autoref{Overview of repair pattern selection} depicts a decision tree used to show repair patterns available to \sysname. The series of dots indicate 
other operations (\textit{e.g.,} subtraction, bit shift, etc.) or arrows not depicted in this figure due to page constraints.
Note that other repair patterns for \texttt{int64\_t}, \texttt{int}, \texttt{short}, and \texttt{unsigned int} types are not depicted in order to not clutter \autoref{Overview of repair pattern selection}. \paul{This tree contains, for example, 
$C1$ up to $C24$ which are the criteria needed by \sysname for choosing a repair pattern.} A repair is selected when at least two criteria are met. \paul{As such, multiple repair patterns may apply for a single repair location in case these satisfy the criteria of multiple patterns.}

For example, the six arrows on the left bottom pointing up depict six repair patterns determined between $C1$ and one of the criteria from $C7-C12$. Depending on whether one of the components of the analyzed statement \texttt{type x = y operator z;} is a constant, a variable or a variable with side effects (\textit{e.g.,} \texttt{int z = i++;} or \texttt{short y = foo();}) a different repair pattern will be proposed.

More specifically, from top to bottom, the decision tree contains an addition (left node) and a multiplication (right node) operation that are used to determine the kind of operation, which is performed in the faulty statement. At the next level in the tree (top to bottom), the type of result variable is determined. Further, at the next level in the tree, it is determined if the parameters of the analyzed statement are constants, variables or have side effects. \autoref{Overview of repair pattern selection} depicts in total 360 repair patterns (360 possibilities to combine two criteria from the range $C1$ to $C12$ with each other, \textit{e.g.,} $C1$ and $C8$, see corresponding arrow in \autoref{Overview of repair pattern selection}) multiplied by five (data types) and the result is then multiplied by two, two main tree branches, \textit{e.g.,} \texttt{add} and \texttt{multiply}) for the \verb!C! statement \texttt{type x = y operand z;}.

\begin{table}[ht!]
\includegraphics[width=90mm]{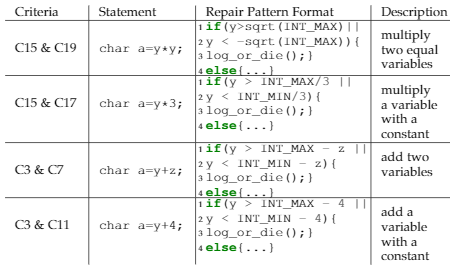}
\caption{Four repair patterns of \sysname.}
\label{Repair Patterns.}
\end{table}

\autoref{Repair Patterns.} depicts four notable repair patterns in more detail.  
We omit the contents of the \textit{if} and \textit{else} branches in order to provide simple to understand repairs.
The \textit{if} branch contains the code to log the occurrence of an integer overflow or 
the code which terminates the execution of the program\footnote{Note that terminating the program is an option 
that is not safe in critical 
program executions, \textit{e.g.,} in case monetary transactions depend on the program execution 
or when the program is driving a train, etc.}.
The \textit{else} branch contains
a statement where the integer overflow was detected. 

Consider the \verb!C! statement $544.$ $a = b * b;$ where an integer overflow was detected at line number 544 in a program's source code file. 
In this case, the pattern depicted in \autoref{Repair Patterns.} on row one ($C15$ \& $C19$) will be used to avoid the fault. 
This pattern will be selected based on the fact that two equal 
variables are multiplied. 
As a result, the statement will be surrounded with the code of the pattern in which
parts are replaced based on the type of statement, 
\texttt{$543. \ if ( ( b > 0\ \&\& \ b >= sqrt(INT\_MAX)) \ || \ ( b < 0 \ \&\& \ b < -sqrt(INT\_MAX) ) )\{545. \ log\_or\_die();\}  546. \ else \{\ a = b * b;\}$}. 
Note that the numbers ($543$ up to $546$) used in this example represent the source code line numbers of the repaired program, respectively.

\emph{How to  construct new repair patterns?}
The repair patterns are constructed programmatically in Java. More precisely, each repair pattern is contained in a Java method which implements a template where the template meta-variables are meant to be filled with \verb!C! code (\texttt{if} and \texttt{else} statements and program variables) which are identified during fault detection. The programmer interleaves the fixed, generic part of the code with the parts which are dependent on each specific detected overflow.

\emph{How were the repair patterns designed?}
We first manually analyzed the Juliet test-suite for identifying the relevant repair patterns. Essentially, each code repair pattern is constructed by first analyzing the \verb!C! source code statement in which an overflow could appear. Then, we identify a program branch condition which repairs the overflow if placed in such a way that it encloses the faulty statement. The goal of the \texttt{if} based condition is to avoid the specific computation which could lead to an overflow in the potentially faulty statement.

{\emph{How does a repair help to fix a fault?}} A generated repair helps to fix a fault by providing to the programmer a finished repair that contains complex mathematical constraints, which are not always obvious.

{\emph{Why are the produced repairs correct?}} A produced repair is correct by construction w.r.t. to the repair correctness Definition \ref{def2}. On the implementation side, after repair insertion, the whole program together with the repair is recompiled; this discard potential faults introduced by the repair.

\paul{\emph{Are the repair patterns complete?}
To our experience, the union of all repair patterns covers all relevant cases we have listed. However, we make no claim about formal completeness, and there may exist corner cases that are not covered by the patterns.
}

Lastly, note that building a repair is not straightforward, as the construction relies on complex mathematical constraints that need to be inferred based on the currently detected fault. Next, these constrains need to be carefully plugged into each repair pattern in order to provide the final repair.

\subsection{Build Repair}
\label{sec:repair}

The automated repair generation algorithm of \sysname consists of the following steps.
 Determine the integer upper bound value (Step 1: \autoref{step1});
 Generate an SMT constraint system (Step 2: \autoref{step2});
 Select constraint values (Step 3: \autoref{step3});
 Recompute the bound checking constraints (Step 4: \autoref{step4});
 Determine the fault type (Step 5: \autoref{step5});
 Select the repair pattern (Step 6: \autoref{step6});
 Determine the new constraint SMT system (Step 7: \autoref{step7});
 Generate code repair (Step 8: \autoref{step8}).

This algorithm corresponds to the box \ding{185} depicted in \autoref{Overview of IntRepair}. The generated repairs are valid and correct according to
Definition~\ref{def1} and Definition~\ref{def2}.
These eight steps are performed in consecutive order as highlighted next.
\subsubsection{Step 1. Determine Integer Upper Bound Value}
\label{step1}
\textsc{}
In order to determine the currently used integer upper bound value in the analyzed program, \sysname performs the following procedure, which allows \sysname to be a multi-precision tool in the sense that it automatically  determines the integer precision needed for each analyzed program.
First, \sysname retrieves the hardware overflow limits\footnote{\textit{E.g.,} by parsing  \texttt{/usr/include/limits.h} on Linux OS.}
for each integer type.
Second, during the symbolic execution-based traversal of the analyzed program path, \sysname searches for previously defined 
and used integer upper bound program variables.
This search is realized by comparing each declared or used variable name (\textit{e.g.,} \texttt{data} in the code snippet depicted in \autoref{Integer overflow based memory corruption example.}) contained in the 
currently analyzed program execution path with one of the supported
integer upper bound values (\textit{e.g.,} \texttt{CHAR\_MAX}, \texttt{INT\_MAX}, \texttt{LLONG\_MAX}, \texttt{SHORT\_MAX}, and \texttt{UINT\_MAX}). Third, in case such an upper bound value is found, it will be set to be the currently used integer upper bound value. 
\sysname can automatically detect for each environment (system), in which it runs, the currently used 
integer overflow upper bound limit (\textit{i.e.,} \texttt{INT\_MAX}) value.
This way, the integer precision is determined for each analyzed program individually. 
Next, this upper bound value will be used to validate the generated candidate code repairs and also to search for integer overflows.

\subsubsection{Step 2. Generate an SMT Constraint System}
\label{step2}
Next, the symbolic variables and the constraints used 
inside the integer overflow checker
are grouped together and stored for later processing.
In particular, \sysname stores:
(1) the statement where an integer overflow is detected,
(2) the SMT formula used to detect the fault in the first place,
(3) the fault ID of the integer overflow checker which was used to detect the fault,
(4) the symbolic variable which was used to detect the integer overflow, and 
(5) other symbolic variables on which the integer overflow triggering variable may depend.

%

For instance, let us consider 
a \verb!C! language assignment statement \texttt{int result} = \texttt{varA} + \texttt{varB;}
and its SMT counterpart \texttt{(assert} \texttt{(= resSymbolic (} \texttt{+ varAsymbolic} \texttt{varBsymbolic)))}.
The \verb!C! assignment statement depicts the addition of two variables and the result is stored in a third variable, with a potential integer overflow fault. 
For this statement, the information collected by \sysname is:
(1) \texttt{int result} = \texttt{varA + varB;},
(2) \texttt{... (assert (}\texttt{= resSymbolic (} + \texttt{varAsymbolic varBsymbolic)));} \texttt{... checksat}, note that this SMT constraint represents a part of the SMT system used to detect the fault,
(3) \texttt{ID-Integer\_Overflow\_Fault},
(4) \texttt{resSymbolic}, and
(5) the \texttt{varAsymbolic} variable.

\subsubsection{Step 3. Select Constraint Values}
\label{step3}
Next, \sysname selects the relevant SMT constraint variables 
based on the type of \verb!C! language statement where the integer overflow is detected.
For example, consider assignment
\texttt{int result} = \texttt{varA +} \texttt{varB;} where the variable \texttt{result} and potentially the variables \texttt{varA} and \texttt{varB} need to be taken into consideration, because these
influence the occurrence of the integer overflow directly.



Further, the symbolic variable, \texttt{resSymbolic}, was
selected by \sysname to be further constrained. This is done in order to check if the code repair that will be generated could remove the previously detected integer overflow fault. For fault removal checking, the whole SMT constraint system slice of this symbolic variable will be used. Note that this constraint system was previously (before running the fault removal analysis) determined during fault localization.
More specifically, this variable will overflow in case of using too large values that cannot properly be stored in the result variable.
Note that depending on the complexity of the analyzed statement, more or fewer variables can
be taken into consideration in order to determine if the previously detected integer overflow would further manifest after re-constraining. This depends on how the 
selected symbolic variables were re-constrained. 
The intuition is that the integer overflow can be detected before a certain variable will overflow. This type of behavior
is useful when generating repairs that are constraining more than one program variable.
Finally, the selected variables will be used in the next step when re-constraining the bounds during the checking of SMT constraints of the SMT system. Note that this SMT system was used upfront to detect the integer overflow. 

\subsubsection{Step 4. Recompute the Bound Checking Constraints}
\label{step4}
In order to solve the problem of determining a suitable variable range for avoiding an overflow,
\sysname applies a specific technique for re-constraining the symbolic variable which has overflowed and
which was selected in the previous step.
The variable can be re-constrained after collecting the program execution path constraints for a single program execution path, 
which were used to check for the presence of an
integer overflow. The presence of an integer overflow fault is indicated if for the selected SMT system the Z3 solver reports \texttt{SAT} 
(satisfiable, integer overflow fault present).


%
\sysname re-constraints (for example, through integer upper bound negation) the variable(s) selected from the previous 
step: the original SMT equation,
\texttt{(assert (>}  \texttt{resSymbolic  2147483647 ))} and the
negated SMT equation,
\texttt{(assert (<=}  \texttt{resSymbolic  2147483647},
in order to determine a potentially safe interval which will not lead to a second integer overflow of the symbolic variable.
Note that other iterative techniques are possible (\textit{e.g.,} iterating backwards through a vector (from large to small values) of consecutive large 
values and checking by selecting each value once as integer upper bound if the check conditions will hold).
The goal is to determine if there is a second integer overflow if \sysname re-constraints the selected variables as mentioned above.
For this purpose, \sysname will check in the next step if for the new SMT constraint
system 
it gets an \texttt{UNSAT} (unsatisfiable, no integer overflow fault present) solver reply. The new constrained SMT will be composed of the old SMT constraint system, which was used to detect the integer overflow, complemented with the re-constrained SMT equations. 
If \sysname gets an \texttt{UNSAT} solver reply, then it determines that there will be no integer overflow if it re-constraints the selected variable(s) 
with the new constraints (\textit{e.g.,} variable range negation, etc.) and as such the integer overflow can be avoided. 
Finally, the information collected at this step will be used in later steps.

\subsubsection{Step 5. Determine the Fault Type}
\label{step5}
For a unique detected overflow, \sysname attributes a report containing a unique fault identifier. 
Based on the generated fault 
identifier (ID), \sysname can determine which fault type it currently deals with. 
This information is extracted from data stored during step two.
With this ID, \sysname checks in the list of currently supported checkers to which checker this stored information belongs to
and determines which repair patterns can be used to repair 
the previously detected fault. 
In this way, repair patterns usable for other types of 
faults will not be suggested when repairing integer overflows.

\subsubsection{Step 6. Selecting Repair Pattern} 
\label{step6}
\label{Selecting Repair Pattern} 

Based on the previously determined fault identifier (ID) in the previous step, \sysname selects the suitable repairs for this integer overflow from the pool of repair patterns using the decision tree (see \autoref{Overview of repair pattern selection} for more details), as discussed next.


\medskip\noindent\textbf{\textit{Pattern.}} 
The repair patterns of \sysname need to address the following challenges.
Each pattern needs to incorporate complex conditions with multiple 
branches depending on the type of the potentially overflowing code location. 
An \sysname pattern needs to be pre-classifiable for typical code locations (\textit{e.g.,} statements) where it would best apply depending on
the AST of the faulty statement(s).
Further, each pattern needs to have at least two branches: (1) in the 
case the check succeeds, and (2) in case the check does not succeed w.r.t. fault logging. The pattern needs to have several customizable components
which can be altered during static analysis with: context dependent values, mathematical functions, or \verb!C! functions extracted from the faulty statement.
As such, \sysname's repair patterns need to be nontrivial fragments of incomplete code, which have to be versatile 
and applicable to different integer overflow repair scenarios.

\medskip\noindent\textbf{\textit{Repairing.}} 
The repair patterns impose the following challenges on \sysname's repairing process. 
Our technique is fully automated and in case \sysname generates more than one possible repair for a single fault, then the user gets a window shown in which he can decide which repair to apply. This step is optional and can be turned on/off from the configuration file of \sysname. In case this step is turned off, then \sysname, in case of multiple repairs generated, chooses the first generated repair and inserts it in the program. Before repair insertion the repair pattern is selected by \sysname with the help of the decision tree.
Further, the automated repair generation mechanism of \sysname, before insertion, needs to be able (1) to extract the components of the faulty statement such as variables, functions (\textit{e.g.,} \texttt{malloc}, etc.) and reuse them for augmenting the selected pattern; to achieve this, the repairing mechanism needs to be able to match these extracted components with the parts of the pattern where these fit, 
(2) next, the repairing mechanism needs to be able to precisely incorporate the faulty statement(s) inside the newly selected pattern,
(3) the repairing mechanism needs to be able to precisely delete the previous statement(s) where the fault was detected and 
to rewrite the existing source file with the newly generated repair pattern at the correct source code location.
As a consequence, the challenges imposed by the repair patterns and \sysname's repairing process are nontrivial and 
require high precision during source code file rewriting.

\medskip\noindent\textbf{\textit{Repair Pattern Description.}} 
\sysname's repair patterns consist of \verb!C! code skeletons where different repair 
parts will be replaced with:
(1) concrete values after their values have been computed,
(2) mathematical operations (\textit{e.g.,} division by a value, etc.), and
(3) standard \verb!C! library functions.
Depending on the context, placeholder variables
will be replaced with corresponding mathematical functions such as the square 
root function \texttt{sqrt} or other functions. In the situation in which a 
mathematical function is used, \sysname does not need to compute the value of the function 
upfront (during static analysis) but rather leave this to be computed later
during symbolic execution analysis (repair validation) or program runtime. 
This offers the advantage that \sysname does not need to be able to compute any possible mathematical function.

Further, the code repair patterns used by \sysname are based on preconditions similar to IOC's~\cite{ioc} 
checks (see \autoref{Overflow and Underflow Checks} for more details). 
These preconditions can cover different types of mathematical operations such as multiplication of 
numbers and addition of variables.
At the same time, the repair patterns are highly configurable and versatile. For example, the programmer can easily change, if needed, the fault handling function 
inside the repair pattern or can extend the precondition such that it captures more complex fault avoiding preconditions.
This can be achieved by modifying a few lines of code inside an existing pattern or by creating 
a new pattern and defining the conditions 
(\textit{e.g.,} depending on the structure of the AST of the statement 
where the fault was detected) 
when such a pattern fits best.

\medskip\noindent\textbf{\textit{Selecting Repair Patterns.}} 
Next, we describe the steps used for selecting a code repair based on a \verb!C! statement having 
three components, \texttt{leftHandSide}, \texttt{rightHandSide}, and \texttt{operator}. 
This translated into the previous example are: \texttt{int result =} \texttt{varA +} \texttt{varB}.
However, \sysname can deal with more complex statements as well, see \autoref{Overview of repair pattern selection}
for more details.

\sysname follows the following steps to select a suitable repair pattern.
First, the code statement where the integer overflow fault was detected 
is divided into its components based on its AST. For example the AST components 
of a \verb!C! statement such as \texttt{int result = varA + varB;}
are \texttt{leftHandSide = varA}, \texttt{operator=+} and \texttt{rightHandSide = varB}.
Second, a series of rules are checked against the AST of the previous \verb!C! statement as follows: 
(1) is \texttt{leftHandSide} equal or is different than \texttt{rightHandSide}, 
(2) what type of operator do we have in the statement, and
(3) how many components does the statement have after the \texttt{=} sign, and so on.
\paul{Third, based on these rules the repair pattern which satisfies the highest number of constraints determined by the previously mentioned rules will be selected.}
Note that each repair pattern has a list of properties attached to it (\textit{e.g.,} use when \texttt{rightHandSide = leftHandSide} and 
\texttt{operator = +}, etc.) 
 that are checked against the above stated rules. This list of properties is statically 
defined when the repair patterns were manually added to the pool of available repairs inside \sysname.
Further, in case there are more repair patterns that fulfill the same number of rules, \sysname selects the first repair pattern 
occurring in the list. In case \sysname cannot determine the pattern selection criteria due to, for example, a complex \verb!C! language statement, then no repair will be proposed.

Lastly, note that, if needed, this approach can be updated such that all legitimate repair patterns will be proposed and for each a repair can be generated, and selected with a human-in-the-loop approach.

\medskip\noindent\textbf{\textit{Repair Pattern Example.}} 
In the following, we present a repair pattern used by \sysname.

\begin{figure}[ht!]
    \includegraphics[width = 90 mm]{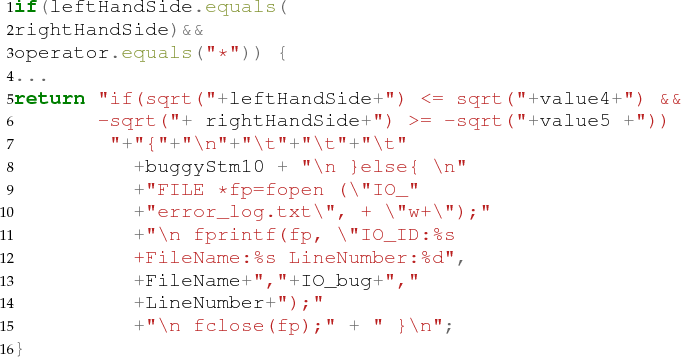}
    \vspace{-.25cm}
    \captionof{figure}{Repair pattern example.}{\unskip}
    \label{Different types of repair patterns.}
\end{figure}
\autoref{Different types of repair patterns.} 
represents a code repair pattern used by \sysname 
during integer overflow fault repairing.
This code repair pattern contains \verb!C! code compatible snippets (shaded in red color) interleaved with several stub variables which will be replaced with concrete variables names, values or mathematical functions depending on the type (depending on its AST structure) of code statement containing the fault.
For example, the repair pattern depicted in \autoref{Different types of repair patterns.} will be used by \sysname when 
the \texttt{leftHandSide = varA} is equal (\textit{e.g.,} string wise comparison) to the \texttt{rightHandSide = varB} and the
\texttt{operator} is equal (\textit{e.g.,} string wise comparison) to the product operator $*$.
After these checks have been performed, the repair will be assembled as presented in the following steps.

First, \texttt{value4}, \texttt{value5} and \texttt{faultyStm10} located at lines 5, 6, and 8 in \autoref{Different types of repair patterns.}, are replaced with:
(1) the squared root of the currently selected integer upper bound value \texttt{value4 $\leftarrow$ sqrt(2147483647)}, 
(2) the negated integer upper bound value \texttt{value5 $\leftarrow$ -sqrt(2147483647)}, and
(3) the program code statement that contains the previously detected integer overflow fault \texttt{faultyStm10 $\leftarrow$ int result = data * data;}. 
Second, the variables \texttt{FileName}, \texttt{IO\_fault} and \texttt{LineNumber} located 
at lines 15-16 in \autoref{Different types of repair patterns.} are replaced with concrete values obtained during fault detection.
Finally, note that: 
(1) other code repair patterns can be selected and used based on the format of the AST of the program faulty statement,
(2) our technique can be easily generalized to more complex \verb!C! code statements than the ones mentioned herein, and 
(3) each generated repair can easily be customized to fit to different types of integer overflow mitigation 
scenarios, \textit{e.g.,} fault logging, etc.

\subsubsection{Step 7. Determine New Constraint System}
\label{step7}

In this step, \sysname assembles the new SMT constraint system which is used to determine if the previously detected integer
overflow is still present. 
During this step, \sysname takes the constraints determined at the previous step \paul{(see \autoref{Selecting Repair Pattern} for more details)}
and inserts them in the 
SMT constraint system which was used to detect the integer overflow. 
%
%
Before inserting the new constraints in the SMT system, \sysname needs to 
\begin{figure*}[ht!]
\centering
\includegraphics[resolution=100000, width=1.9\columnwidth]{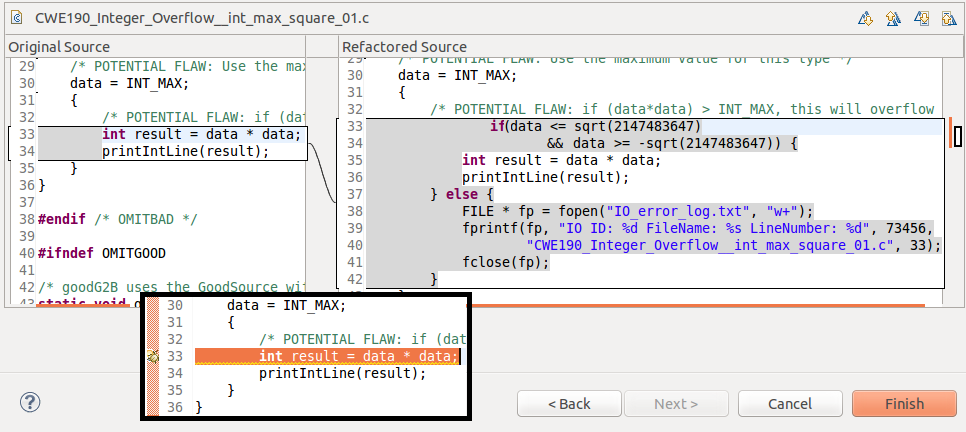}
\caption{\sysname's view of the localized fault (left hand side) and the program repair (right hand side).}
\label{Repair Insertion Steps}
\end{figure*}
remove the original SMT constraints which were used to detect the presence of the integer overflow as follows.
\texttt{(assert (>} \texttt{resSymbolic} \texttt{2147483647 ))} is replaced with:
\texttt{(assert (<} \texttt{resSymbolic}
\texttt{ 2147483647 ))}.
\paul{The decision of removing these SMT constraints that are used to detect the integer overflow is made based on logical inter-dependencies of these SMT statements}.
More precisely, since the SMT constraints used to check for an integer overflow are added in the integer overflow checker 
in an incremental manner, \sysname can precisely determine which constraints can be safely removed and replaced with 
new ones determined at the previous step.
Next, this SMT system will be fed into the Z3 solver. In case the solver replies \texttt{SAT}, then the 
constraints added represent valid constraints which can be 
used to avoid the previously detected integer overflow. 
This new SMT constraint system serves as ground truth w.r.t. the fact that the integer overflow can be removed if certain 
symbolic variables are re-constrained in an appropriate way.
Finally, note that the above-described symbolic variables have concrete 
counterparts values, which will be inserted inside the code repair when assembling it.

\subsubsection{Step 8. Generate Code Repair}
\label{step8}
\label{Generate Code Repair}
This step consists of putting together the final code repair(s) and saving them into a list in case multiple repairs are suggested. 
After the repair components have been inserted into the previously selected code repair pattern, \sysname generates 
a \verb!C! source code repair which is 
syntactically correct, can be compiled and can be further on edited after insertion (if desired). 

\subsection{Implementation}
\label{sec:impl}
We have implemented the approach  presented in~\autoref{sec:design} in 
a prototype named \sysname.
It is an Eclipse ~\cite{cdt} plugin based on Codan~\cite{codan}.
\sysname consists of approx. 10 KLOC and
is developed as an Eclipse plug-in mainly because:
(1) the Eclipse CDT API can be easily reused,
(2) a GUI is easily obtainable, and 
(3) the obtained tool can be used in both online (during code typing) and off-line (after finishing code typing) modes. 
Note that Codan~\cite{codan} is used by \sysname to construct the program CFG,
analyze the program AST statements, and
perform bottom-up traversals by using a \verb!C! program statement visitor in order to construct SMT constraints.
Three additional auxiliary tools are implemented as well to work with \sysname.

The assembled repair generated in \autoref{Generate Code Repair} is sent to the Eclipse language tool kit (LTK) API~\cite{ltk} based component of our engine, which will assemble 
the repaired code.
The LTK component adds information on how to position the repair in the faulty program such that the integer overflow will not occur after the repair was inserted. 
In case multiple repairs have been generated, \sysname
saves these repairs in a list of repair candidates for the previously detected integer overflow.
Lastly, the repair which was selected by the programmer will be passed to the \sysname repair insertion 
component, which will create two differential views, with the repair inserted in the file containing the fault and without.

Next, 
we briefly describe the three additional tools which we implemented.
First, we have a generator  of Eclipse CDT projects which is used for generating Eclipse CDT compatible projects from Juliet cases.
Second, we have a source code refactoring tool based on the Eclipse LTK~\cite{ltk}, JFace~\cite{jface}, and Eclipse CDT~\cite{cdt}, for inserting repairs at the source code level.
Third, we have a specific test case generator to assess the correctness of IntRepair.
\paul{This tool works as follows. First, the assessed Juliet test suite program is manually annotated with strings that localize the fault (with an annotation one code line before the code line which contains the fault). Second, our test case generator tool parses the source code of the faulty programs (our test resources) and searches for this annotation. Third, it generates test cases based on code skeletons which contain, for example, the line number where the true positive fault was detected as well as the file name and function name in which the fault was previously detected. Finally, when a generated test case is executed, it is checked whether a reported fault is located at the line number, in the function and in the file which were previously annotated.}

\subsection{Graphical User Interface in an IDE}
\label{sec:insertion}

\autoref{Repair Insertion Steps} shows the \sysname Graphics User Interface (GUI).
First, the integer overflow triggers a fault marker depicted in the black bordered box with
a yellow fault icon.
It means that on the \verb!C! statement an
integer overflow fault was detected.
Second, by right clicking on this fault marker the user can start the code
re-factoring wizard.
The code re-factoring wizard is composed of two windows.
The first window is used to make repair type decisions (currently only in-place repairs are available). 
The second window depicted in \autoref{Repair Insertion Steps} (in the background)
contains a differential files view visualizing the differences between the original
file containing the fault and the modified file with the selected repair inserted. This second window helps the developer to analyze the repair before it is applied.
Finally, it is possible to navigate between these two windows. In case the programmer decides to insert the generated repair (see~\autoref{sec:repair}), then this can be achieved by left-clicking on the \texttt{Finish} button.

\section{Evaluation}
\label{sec:evaluation}
In this section, we present evaluation results and address the following research questions (RQs):
\begin{itemize}
\item \textbf{RQ1:} How effective is \sysname in repairing integer overflows? To what extent does \sysname produce false positives? To which degree does \sysname ensure that a repair completely removes the previously detected integer overflow?

\item \textbf{RQ2:} What is the performance impact of \sysname? 

\item \textbf{RQ3:} How do the fault repairs of \sysname preserve program correctness? 

\item \textbf{RQ4:} How effective is \sysname compared to manual repair of integer overflows? 
\end{itemize}

\subsection{Evaluation Setup}

\begin{table}[ht!]
\resizebox{\columnwidth}{!}{%
\begin{tabular}{ l | l | r | l}
  Subject Prog.         & LOC       & \# Programs & Description \\ \hline
  CWE-190         & up to 638 & 2,052  & Open source testsuite    \\
  Benchmark       & up to 20K & 50    & We introduce a benchmark \\
                  &           &       & of synthetic programs    \\
  
\end{tabular}}
\caption{Overview of the subject programs.}
\label{subject programs}
\end{table}
\autoref{subject programs} depicts a summary of the programs used to evaluate \sysname. Next, we give details about the used programs.

\textbf{\textit{Juliet benchmark.}} First, we selected 2,052 \verb!C! programs contained in SAMATE's Juliet test suite~\cite{juliet}, which is the largest open 
source test suite for \verb!C/C++! code to the best of our knowledge. 
From the set of all  integer overflow test cases, we kept all 2052 \verb!C! programs, but we excluded the 198 \verb!C++! programs which are out of scope of our work.
In particular, we consider the CWE-190 category. Each program contains on average 
476 LOC with a maximum of 638 LOC. Each program has integer overflows: exactly one true positive and several false positives (see characteristics of the Juliet test suite~\cite{juliet} for more details). We used these programs as they contain all possible program control flows which may lead to an integer overflow. Further, these help to assess previously known faults in a systematic way.

\textbf{\textit{Synthesized Code.}} In addition, we build a benchmark with 50 synthesized programs with complex control flows and with a high number of code lines. 
\paul{We synthesized the programs by first manually selecting five representative seed programs contained in the Juliet test suite. For these original programs, we added additional code in order to increase the number of LOC in order to reach up to 20 KLOC. The added LOC represent chains of function calls and loops which cause the program to loop for numerous times through the code as the program runs. The objective is to trigger the same overflow but in a convoluted manner.}

In each synthesized program, the exact number 
of seeded integer overflows is known. The benchmark allows us to demonstrate that \sysname can scale efficiently to large, and complex programs.
The synthesis of programs is parameterized as follows: 
(1) the total number of function calls,
(2) the number of loop iterations, and
(3) each generated program contains exactly one true positive and 
a variable number of false positives, and the true positive should be located deep inside the program,
\textit{e.g.,} several thousands of branches nested inside the program execution tree.
Using these limits, we generate 50 programs ranging from 6 to 20 KLOC. 
The synthesis algorithm iterates through several loops in which it adds functions in the program
which are calling each other. These functions contain a variable
number of branches which are also counted until a certain depth is reached.

\textsc{}
\begin{table}[ht!]
\begin{tabular}{ p{1.2cm} | p{1.5cm} | p{1.5cm} | p{1.2cm} | p{1.2cm}}
  Subject Prog.        &Satisfiable Paths &Unsatisfiable Paths &Program Branch Nodes & Program Branches\\ \hline
  CWE-190        & 115 K.           &1.2 Mil.            &5.73 Mil.            & 12  Mil.         \\
  Benchmark      & 45 K.            &0.7 Mil.            &4.4 Mil.             & 8.8 Mil.         \\
\end{tabular}
\caption{Descriptive statistics of our subject programs.}
\label{Analyzed program statistics.}
\end{table}
\textbf{\textit{Statistics.}} 
\autoref{Analyzed program statistics.} depicts several characteristics
of the selected 2,052 analyzed programs contained in SAMATE's
Juliet test suite and the 50 programs of our benchmark. 
The number of satisfiable program paths refers to paths for which the symbolic analysis component of \sysname was able to reach the target fault. An unsatisfiable path is obtained when the symbolic analysis of \sysname gets an unsatisfiable reply from the SMT solver.  \footnote{The number of paths for the 50 synthesized programs of the benchmark is smaller compared to the 2,052 \texttt{C} programs because we focused on adding loops and not on adding branches. For this reason, it makes no sense to compare those two numbers.}
Further, \autoref{Analyzed program statistics.} depicts the number of analyzed program paths which were extracted from the program control flow graph. Next, in \autoref{Analyzed program statistics.}, we report the total number of branch nodes as well as program branches (C branch statements) along all program paths which \sysname has analyzed. \autoref{Analyzed program statistics.} shows that \sysname can scale to large programs.

The programs contained in the CWE-190 contained on average:
56 satisfiable paths,
584 unsatisfiable program execution paths, and
5,800 program branches, which were counted during
program analysis. In contrast, the programs contained in our own benchmark
contained on average:
900 satisfiable paths,
14,000 unsatisfiable program execution paths, and
28,000 program branches, which were counted during the analysis of the programs.

\textbf{\textit{Note.}}
Unfortunately, we cannot compare \sysname with any existing approach as there are no open implementations which can statically analyze source code, detect the integer overflow and provide a repair at the source code level. Also, we are not aware of any static \verb!C! source code analysis tool which can be used to run on the same programs as \sysname.

\paul{\textbf{\textit{Test case generation}}. The test cases were generated automatically by a tool which was specifically created for this purpose, described in \autoref{sec:impl}.}

\textbf{\textit{User Study.}} We also performed a controlled experiment with 30 participants in which we assessed the efficiency of \sysname 
during integer overflow repair. The used protocol is described in \autoref{Controlled Experiment (RQ5)}.

\textbf{\textit{Experiment Setup.}} 
Experiments were conducted on a Dell desktop with an Intel CPU Q9550 \url{@} 2.83GHz, 64-bit, 12GB RAM, Eclipse IDE Kepler and OpenSuse 13.1 OS.

\subsection{Effectiveness (RQ1)}
\subsubsection{Fault Detection}
In this section, we show the effectiveness of \sysname w.r.t. fault detection. As our main test-bed, we select 2,052 \verb!C! programs contained in SAMATE's Juliet test suite~\cite{juliet} and 50 programs from our customized benchmark.

\textbf{\textit{Methodology.}}
We prepare the 2,052 \verb!C! programs in Juliet in order to be handled by \sysname.
Further, we generate 2,052 test cases which are used to dynamically assess 
if the detected integer overflow faults are detected at the 
correct source code location. For the automated generation of these test cases \paul{we used the automated test case generation tool presented in \autoref{sec:impl}.}
We conduct the same pre-processing with the 50 programs from our own benchmark.

(1) We assess the effectiveness of integer overflow detection by running 
the generated test cases for all programs (including the 2,052 \verb!C! programs in the Juliet test suite and 50 \verb!C! programs contained in our benchmark) and checking the generated reports.
More precisely, we checked whether each report describes an integer overflow reported at the expected location.
(2) For those reports which are confirmed as true positives, \sysname inserts a suitable repair into the overflow site. 
(3) Then, \sysname was run again in order to check if this was a true positive and if it was removed. 

We point out that, for the assessed programs, \sysname did not detect any false positives.
Further, \sysname is able to detect and repair all previously detected true positives present in the analyzed programs without repairing any false positives. 
Finally, note that the ground truth is that
each of the 2,052 \verb!C! programs contains one genuine integer overflow (if detected this generates a true positive)
and multiple not genuine overflows (if detected these generate false positives) and the same is true for the
programs contained in our benchmark.


\begin{mdframed}[style=MyFrame]
\textbf{\textit{Results.}}
According to our experiments, \sysname is able to detect all integer overflows (2,052 overflows contained in the Juliet test suite and 50 in our benchmark) at the expected locations in the analyzed programs. 
After the repairs conducted by \sysname, the integer overflow reports are no longer generated. This means that \sysname is able to perfectly repair all integer overflows for the analyzed programs.
\end{mdframed}

\subsubsection{Fault Removal}
\label{Fault Removal}
In the section, we explain how the integer overflows are completely removed by \sysname.
Recall that \sysname checks if a fault was completely removed after a repair was inserted in two operation modes.

\textbf{\textit{Methodology.}} In this experiment, we use \sysname's \textit{Manual Mode}.
In this mode, the user can again analyze the repaired program by manually restarting the static analysis. In case the symbolic analysis detects no other (or the previously fixed) integer overflow, then no fault report will 
be generated. This helps the programmer to conclude that the fault was completely removed and no new fault 
was inserted into the program. 
Indeed, we used \sysname to re-analyze each repaired program. For each of the repaired programs, we checked if after the program was repaired any new fault report was filed. 

\begin{mdframed}[style=MyFrame]
\textbf{\textit{Results.}} After repairing and subsequently reanalyzing each program, no new overflow reports are generated, meaning that the faults are completely removed without inserting new ones.
For all programs, the faults
are successfully removed by inserting the repair at the
correct location.
\end{mdframed}

\subsection{Performance (RQ2)}
In this section, we evaluate the performance of \sysname in three ways
(1) the time \sysname takes; 
(2) binary size blow-up, and 
(3) the runtime overhead caused by the overflow repair.
Finally, we show that \sysname has the potential 
to scale to large programs as well.

\subsubsection{Static Analysis Time}

The time that \sysname spends in static analysis, \textit{i.e.,} static symbolic execution and overflow repair processes, is an important criterion to assess the usability of our tool.

\begin{table}[ht!]
\centering
\begin{tabular}{ r | r}
 6 KLOC Programs    &46 sec. ($\textless$ 1 Min.)  \\
11 KLOC Programs    &151 sec. ($\textless$ 3 Min.) \\
20 KLOC Programs    &567 sec. ($\textless$ 10 Min.) \\ 
\end{tabular}
\caption{Average repair generation time in seconds.}
\label{Analyzed program statistics time.}
\end{table}

\begin{mdframed}[style=MyFrame]
\textbf{\textit{Results.}}  \autoref{Analyzed program statistics time.} depicts the average execution time over 10
runs for each of the benchmark programs grouped in three
main categories based on their LOC. \sysname handles all programs under 10 Min. (567 sec.) on average. 
\paul{Consequently, we consider  \sysname to be time-effective and usable in a development environment.}
\end{mdframed}


\subsubsection{Binary Size Blow-up.}
We assessed the source code and binary blow-up by counting the increase in source code lines and in bytes for the programs before and after applying the repairs.

First, we compared the total number of lines contained in the source code against the number of lines of code which were added after inserting all the repairs into the 2,052 vulnerable programs. 
As already mentioned, the initial number of lines was 977.7 KLOC. After applying all repairs, we added in total around 10,045 LOC. 

\begin{mdframed}[style=MyFrame]
\textbf{\textit{Results.}} \sysname yields a binary blow-up of less than 0.56\% in LOC. In our opinion, this represents an acceptable source code lines increase. 
\end{mdframed}

\subsubsection{Execution Overhead.}
We evaluated the runtime overhead introduced by \sysname by comparing the execution time of un-repaired programs against the one of the repaired programs. \paul{This is made possible since each analyzed Juliet program has its own input workload coming from the official Juliet test suite. We use this workload as is in our experiment. For the 50 synthesized programs, we use the input workload of the corresponding seed Juliet program (recall that we use five seed programs from which we generated our synthetic programs.)}

\begin{mdframed}[style=MyFrame]
\textbf{\textit{Results.}} In our experiments, we recorded an average runtime overhead of 1.57\%.
Thus, the inserted repairs do not considerably influence the runtime overhead of the repaired programs. 
\end{mdframed}


\subsection{Correctness (RQ3)}
We verify the correctness of our repairs by checking that our symbolic execution component 
generates semantically different code. In other words,
we want to find out if the repairs influence the program behavior and whether the repair correctly removes the fault (see repair correctness Definition \ref{def2}).

To achieve this goal, we compare the generated SMT model before and after the repair. 
For this, \sysname's symbolic execution component stores the SMT system which was used to trigger
the fault report. Next, this constraint system is compared with the new SMT system which was generated after repair validation. This results in the following analysis process:

\begin{enumerate}
 \item For all programs contained in our benchmark and the Juliet test suite, we store the SMT system for each program and the 
 node, containing source code line number and file name, where the fault is located. 
 \item We apply the repair to the previously detected integer overflow by automatically rewriting the program.
 \item We re-run the symbolic analysis on the repaired program and store the new SMT system.
 \item 
 We semantically 
 checked each SMT system stored at step (2) and each SMT system collected at step (4) based on string comparison of the node with another node, where a node is a 
 complete statement, \textit{e.g.,} \texttt{x = p + 5}. This enables \sysname to detect the differences between these two systems.
 \item We compare the SMT system differences and proceed as follows:
 (1) if the new SMT system only reflects the semantics of the inserted repair: we report no program correctness violation,  
 (2) otherwise, we report that the program correctness will potentially be affected by inserting this repair.
\end{enumerate}

We automatically compared in total 225K SMT models and all repaired programs. These differ by up to five AST nodes and by less than three SMT constraints on average, respectively. 
Further, all checked SMT models differ only with respect to the repair 
semantics and no additional program semantics were introduced in the program due to the repairs.

Additionally, we performed the following evaluation process in order to increase the level of confidence w.r.t. repair correctness.


\textbf{\textit{Recheck Repaired Program.}} We re-run the symbolic analysis, after applying the repair, 
on each of the programs in order to see if there is a
potential new integer overflow fault or the old one is still
present in the repaired program. 

\textbf{\textit{Check Syntactical Program Correctness.}} We recompile (we use the GCC compiler) each of
the programs to investigate if the repaired program is compilable 
and thus syntactically correct. The compiler reported no syntactical errors.


\begin{mdframed}[style=MyFrame]
\textbf{\textit{Results.}} The normal behavior
of the repaired programs does not change after applying the repairs.
\end{mdframed}

\subsection{User Study (RQ4)}
\label{Controlled Experiment (RQ5)}
In order to evaluate \sysname's effectiveness w.r.t. integer overflow repairing compared to manual repairing, we performed a user study 
with 30 graduate students (18 males and 12 females).
We asked all potential participants about their programming experience, and we selected only the ones which had a \verb!C/C++! programming experience. Participants had on average between 1-3 years of industry programming experience.

\textbf{\textit{Methodology.}} 
The experiment was conducted with each participant individually placed at a single computer with an additional person in the room who overlooked the whole experiment. In this way we avoided bias towards the fact that if several participants were in the same room at the same time then they could influence each other.
The computer used in our experiment was equipped with two versions of Eclipse CDT. One computer had an Eclipse version with \sysname installed and the other computer had an Eclipse version without \sysname installed.
In order to avoid gender bias, we split the 30 participants evenly in two groups in such a way that the number of females and males was split evenly between the two groups.
We \textit{randomly} selected in total two programs from our benchmark, and one program from the Juliet test suite. Next, we told each participant, of both groups, to search for a single integer overflow in each of the three programs and to repair it. We instructed each participant that the usage of Eclipse IDE is recommended.

The participants were allowed to read and execute the program. 
Further, all participants had similar experience with the task at hand, particularly fault detection, as they had 1-3 years of experience as developers in different software companies. Lastly, in the case of manual debugging no test cases were prepared for simplicity reasons. The experiment supervisor assessed each fault detected by the participants by manual inspection after it was reported. In order to avoid bias, the participant was not given any feedback on the reported bug.

\textbf{\textit{Group 1.}}
Each participant had access to the latest Eclipse CDT IDE and to the GCC (v. 4.9.3) compiler~\cite{gcc} through terminal access. The Eclipse CDT IDE in this group were installed without \sysname.

\textbf{\textit{Group 2.}}
Before the experiment, each participant from this group watched a short one minute demo movie showing how to use \sysname.
Each participant could search for faults and repair them with the help of \sysname. In this way, we avoided bias related to how fast or slow a participant adapts to the new tool to use during the fault detection and repair task.

We measured the time needed for each participant to locate the fault and repair it and the success rate for each analyzed program after the participant decided that he/she was finished with all three programs at the end of their analysis.
After the experiment, we asked each participant if (1) \textit{he/she would reuse \sysname in his/her daily routine} and if (2) \textit{he/she would recommend
it to other peers}. Each question could be answered with \textit{yes} or \textit{no}.

First, in total, the participants of \textit{Group 1} needed 4,020 seconds (64 Min.) to repair the faults in the experiment, 
whereas participants from \textit{Group 2} only needed 348 seconds (5.8 Min.) to repair them. 
That means, programmers using {\sysname} in our experiment were over 11 times ($4,020/348$) faster compared to relying on debugging and manual repairing.

Second, we inspected the repairs inserted manually by \textit{Group 1} and those inserted by \sysname.
 38\% of the manual repairs were correct, \textit{i.e.,} the integer overflow faults were removed and 
no further vulnerabilities were introduced.
By construction, all the repairs with the help of \sysname were correct (see \textit{RQ2} for more details).
Hence we conclude that \sysname is more effective w.r.t. overflow repair than manual analysis.

Third, out of the total participants of \textit{Group 2}, 80\% (12 participants) found \sysname to be very useful and 93\% (14 participants) would further recommend \sysname to their peers. 
We informally asked these participants about their experiences with \sysname after the experiment. First, the participants provided positive feedback about the usability of the user-interface. Second, having used advanced software engineering tools and/or Eclipse IDE in the past provided an immediate feeling of familiarity with \sysname's concepts and likely also contributed to better task performance.

Lastly, a significant amount of time was needed by the participants in \textit{Group 1} to determine the correct form of the repair. This comprised writing and ordering the conditions, and determining which variables should be best used to properly repair the overflow. Further, the participants reported that they spent a lot of time (around 35\%) on evaluating if the repair is correct and if it really removes the fault.

\begin{mdframed}[style=MyFrame]
\textbf{\textit{Results.}} 
This user study shows that \sysname is usable and improves the repair of integer overflows compared to traditional program repairing which relies mostly on debuggers.
\end{mdframed}

\section{Discussion}
\label{Discussion}
In this section, we discuss the threats to validity of our experiments and the limitations of \sysname.

\subsection{Threats to Validity}

\textbf{\textit{Internal validity}}
depends on the correctness of our prototype implementation. It may also be affected
by the evaluation setting and the execution of the conducted experiments. 
To mitigate the concern about runtime performance validity, we carefully repeated the experiments
ten times and took the average performance values. Further, to validate the correctness
of our automatically generated repairs, we inspected once each program repair manually.

\medskip\noindent\textbf{\textit{External validity}} threats may derive from the selection 
of programs which we used in our evaluation. We validated our repairs on 
2,052 programs and also on seeded programs from our benchmark.
This benchmark has been especially designed for mitigating this threat. 

\medskip\noindent\textbf{\textit{Construct validity}} threats may appear from the fact that our repairs may 
introduce overhead or unwanted program behaviors. We point out that our repairs are suggested 
 only for code locations where an integer overflow was previously confirmed. 
Further, our experiments show that our repairs do not harm performance in a  significant way, do not change an observable
program behavior.

\subsection{Limitations}
\textbf{\textit{Restricted C Language Semantics.}} \paul{IntRepair covers all program constructs contained in the selected Juliet programs.}
Yet, the used static symbolic analysis framework (Codan) only supports a subset of the \verb!C/C++! programming language semantics, which means that full-fledged \verb!C! code cannot be handled. Nevertheless, we do not consider this to be a major scientific limitation, adding all \verb!C/C++! statement types and function headers can eliminate this with sufficient engineering manpower. 
\sysname currently cannot handle all \verb!C! language semantics and as such we cannot run it on \verb!C! programs. To our knowledge, no other open source detection and repair tool exists that can handle the full semantics of the \verb!C! programming language. For this reason, it is also quite difficult to compare our tool with others. Further, in this paper we propose an original idea with a proof-of-concept implementation. Future efforts towards commercial usage of the technique will make it work for arbitrary programs.


\medskip\noindent\textbf{\textit{Complex C Language Constructs.}} 
Complex \verb!C! program expressions cannot be handled due to limited support in the currently available open source state-of-the-art SMT solvers for non-linear computations. 
Thus, we want to address this potential limitation by carefully providing SMT constraints which closely capture the original program constraints and by experimenting with different SMT solvers.

\medskip\noindent\textbf{\textit{Loops and Recursion.}} 
\sysname's implementation depends on program loop and recursion depth unrolling which incurs well-known precision penalties for all symbolic execution-based techniques, and these operations may, as a result, introduce false positives. Note that \sysname does not produce false positives by design as each detected fault is a genuine one. Our static analysis is time-consuming. Its accuracy and performance will affect \sysname's results. \sysname does not have access to runtime information and a real environment is also not available. In order to deal with this limitation, \sysname uses function stubs which can help to simulate the environment (interaction with third-party libraries).
Note that loop unrolling and recursion depth have to be bounded when using symbolic execution-based approaches in general since we want to avoid the path explosion problem and thus infinite analysis time. 
We want to address this potential insufficiency by a prior analysis of program loops in order to derive loop invariants. Further, we plan to extend the list of currently supported stub functions for the main \verb!C! most used system libraries.

\medskip\noindent\textbf{\textit{Path Explosion Issues.}} 
\sysname relies on static symbolic analysis of source code. It has several limitations which are common to all static analysis techniques when comparing to dynamic analysis techniques such as the well-known path explosion problem. Note that static searching through the program paths is different from dynamic path analysis where program path coverage is driven by the provided program input.
We addressed this issue in \sysname by implementing  different path exploration techniques in order to analyze preferentially more fault-prone paths. 

\medskip\noindent\textbf{\textit{Test Programs.}} 
\sysname's evaluation is based on the currently largest open source test suite for \verb!C/C++! programs, \textit{i.e.,} SAMATE's Juliet test suite, which contains complex control flows and a large number of situations where integer overflow can occur.
The programs contained in our benchmark are comparable with large and complex programs.
As a result, the findings of our evaluation do not necessarily reflect the behavior of \sysname when applied to 
larger programs. However, we do not think that this limits the applicability of \sysname, since our 
tool is highly scalable due to its configurable analysis. Further, we want to evaluate \sysname on even larger programs as well.

\medskip\noindent\textbf{\textit{False positives/negatives.}} 
\paul{While comparable with real-world programs, the test programs for \sysname are relatively compact programs. Within this context, we have not identified any false positives or false negatives. However, due to the complexity of the task at hand, we expect that in even more complex and large real-world programs, \sysname would report false positives.}

\medskip\noindent\textbf{\textit{User Study Results.}} 
Lastly, we tested \sysname in a controlled experiment with a restricted number of participants. For this reason our findings may differ from industrial settings where real development conditions are available. \paul{One limiting aspect is that not all of the participants may have used Eclipse in the past for developing in industry.}
Nonetheless, we expect that our tool can help to drastically cut down the time needed for fault detection and repair due to its usability and low intrusiveness. In future work, after implementing all \verb!C! language specific semantics needed for real-world programs, we plan to evaluate \sysname also in industry settings as well.


\subsection{Repair Pattern Generalization}
The repair patterns used by \sysname generalize beyond the programs used in the benchmarks for the following main reasons. First, the decision tree used to select a repair is general enough and its nodes can be extended with other operators and variable types such that for other types of \verb!C! language statements suited repairs can be inferred. Second, the repair patters can be manually extended and customized very easily towards other needs or specific programs. Our repair patterns are designed such that the programmer can tailor the repairs towards his particular needs by updating the repair selection tree and the list of available repairs. Lastly, this makes our repair patterns flexible, customizable and usable for different programs.

\section{Related Work}
\label{sec:relatedwork}
There is a large body of research work focusing on integer overflow \textit{detection}:
Senx~\cite{senx},
ARCHER~\cite{archer},                
UQBTng~\cite{uqbtng},                  
PREfast~\cite{prefast},               
Rich~\cite{rich},                      
SAGE~\cite{sage},                      
CBMC~\cite{cbmc},                      
IntScope\cite{intscope},              
Brick~\cite{brick},                    
IntFinder~\cite{intfinder},            
SmartFuzz~\cite{smartfuzz},             
PREfix~\cite{prefix},                  
IntPatch~\cite{intpatch},              
IOC~\cite{ioc},                        
IntFlow~\cite{intflow},                
SoupInt~\cite{soupint},               
SIFT~\cite{sift},                      
TAP~\cite{tap},                        
Diode~\cite{diode},                  
Indio~\cite{indio},                    
Zhang \textit{et al.}~\cite{toolx}, and              
IntEQ~\cite{inteq}. 

In contrast, only few approaches focus explicitly on integer overflow \textit{repairs}: CIntFix \cite{CIntFix}, IntPTI \cite{intpti}, SoupInt~\cite{sift}, CodePhage~\cite{codephage}, TAP~\cite{tap}, and SIFT~\cite{sift}.
Out of these approaches, only TAP \cite{tap} (technical report) and SIFT~\cite{sift} first explicitly detect the integer overflow and then repair it. The other tools---which do not first confirm  
fault existence---mostly blindly change the code in all fault prone locations in the hope to avoid integer related problems. 
For example: (1) CIntFix \cite{CIntFix} utilizes integers of infinite size with two's 
complement encoding in place of original bounded integers, and (2) AIC/CIT/RAO~\cite{coker:integer} relies on
three code transformations: add integer cast (AIC), replace arithmetic operator (RAO), and change integer type (CIT) to change the program in order to avoid integer overflows.
However, these tools have a high runtime overhead.


SIFT~\cite{sift} first detects the integer overflow (fault detection is based on Diode \cite{diode}) and then generates 
an input filter to eliminate the fault at the binary level. SIFT is a static input filter generation
tool, which inserts input filters in the program binary for which
the source code was previously manually annotated with source code annotations.
The main drawback of SIFT is that it relies on tedious user source code annotations. 

In the above-mentioned approaches, fault localization is performed before repair generation.
In contrast, we combine fault localization and repair generation, and as a result, 
we obtain the capability of generating precise repairs which remove the fault for 
various program inputs (fault detection is not program input independent) and at the exact 
location where the fault was detected upfront. For this purpose, we use SMT solving for fault localization. However,
unlike \sysname, SMT solving is not used for repair synthesis by TAP or SIFT. As such, their fault removal 
process is bound to a limited number of test inputs to confirm that the fault was removed after 
the repair was inserted, which does not guarantee that the fault was really removed for all possible program inputs.

\section{Conclusion}
\label{sec:conclusion}
We presented \sysname, a novel framework which provides comprehensive integer overflow detection, correct repair generation, and validation.
To the best of our knowledge, we provide the first static symbolic execution-based technique
that combines detection, generation and validation of source code repairs for \verb!C! programs.
We applied \sysname to 2,052 \verb!C! programs (approx. 1 million LOC) and to 50 programs contained in our seeded benchmark, which includes programs that have up to 20 KLOC.
Our experimental results show that \sysname was able to effectively detect integer overflows and successfully repair them 
with source code repairs. 
Lastly, our controlled experiment shows that \sysname is over 10 times more time-effective and has a higher repair success rate than manual repairs. 


\section*{Acknowledgments}
The authors are grateful to the anonymous reviewers for
their insightful and constructive comments. Further, we want to especially thank Nenad Medvidović and Víctor A. Braberman for their constructive feedback throughout the reviewing process of this paper.

\bibliographystyle{myIEEEtranS}
 \bibliography{sample-sigconf} 

\begin{thebibliography}{10}
\providecommand{\url}[1]{#1}
\csname url@samestyle\endcsname
\providecommand{\newblock}{\relax}
\providecommand{\bibinfo}[2]{#2}
\providecommand{\BIBentrySTDinterwordspacing}{\spaceskip=0pt\relax}
\providecommand{\BIBentryALTinterwordstretchfactor}{4}
\providecommand{\BIBentryALTinterwordspacing}{\spaceskip=\fontdimen2\font plus
\BIBentryALTinterwordstretchfactor\fontdimen3\font minus
  \fontdimen4\font\relax}
\providecommand{\BIBforeignlanguage}[2]{{%
\expandafter\ifx\csname l@#1\endcsname\relax
\typeout{** WARNING: IEEEtranS.bst: No hyphenation pattern has been}%
\typeout{** loaded for the language `#1'. Using the pattern for}%
\typeout{** the default language instead.}%
\else
\language=\csname l@#1\endcsname
\fi
#2}}
\providecommand{\BIBdecl}{\relax}
\BIBdecl

\bibitem{avgerinos2014enhancing}
T.~Avgerinos, A.~Rebert, S.~K. Cha, and D.~Brumley, ``{Enhancing Symbolic
  Execution with Veritesting},'' in \emph{Proceedings of the International
  Conference on Software Engineering (ICSE)}.\hskip 1em plus 0.5em minus
  0.4em\relax IEEE/ACM, 2014, pp. 1083--1094.

\bibitem{smtlib}
\BIBentryALTinterwordspacing
C.~Barrett, A.~Stump, and C.~Tinelli. (2010) {The {SMT-LIB} Standard Version
  2.0}. [Online]. Available:
  \url{http://smtlib.cs.uiowa.edu/papers/smt-lib-reference-v2.0-r10.12.21.pdf}
\BIBentrySTDinterwordspacing

\bibitem{rich}
D.~Brumley, T.~Chiueh, R.~Johnson, H.~Lin, and D.~Song, ``{RICH: Automatically
  Protecting Against Integer-based Vulnerabilities},'' in \emph{Proceedings of
  the Network and Distributed System Security Symposium (NDSS)}, 2007.

\bibitem{klee}
C.~Cadar, D.~Dunbar, and D.~Engler, ``{KLEE: Unassisted and Automatic
  Generation of High-Coverage Tests for Complex Systems Programs},'' in
  \emph{USENIX Symposium on Operating Systems Design and Implementation
  (OSDI)}, 2008, pp. 209--224.

\bibitem{cadar2013symbolic}
C.~Cadar and K.~Sen, ``{Symbolic Execution for Software Testing: Three Decades
  Later},'' \emph{Communications of the ACM}, vol.~56, no.~2, pp. 82--90, 2013.

\bibitem{intfinder}
P.~Chen, H.~Han, Y.~Wang, X.~Shen, X.~Yin, B.~Mao, and L.~Xie, ``{IntFinder:
  Automatically Detecting Integer Bugs in x86 Binary Program},'' in
  \emph{Proceedings of the International Conference on Information and
  Communications Security (ICICS)}, 2009.

\bibitem{brick}
P.~Chen, Y.~Wang, and Z.~Xin, ``{Brick: A Binary Tool for Run-time Detecting
  and Locating Integer-based Vulnerability},'' in \emph{Proceedings of the
  International Conference on Availability, Reliability and Security (ARES)},
  2009.

\bibitem{CIntFix}
X.~Cheng, M.~Zhou, X.~Song, M.~Gu, and J.~Sun, ``{Automatic Fix for {C} Integer
  Errors by Precision Improvement},'' in \emph{Proceedings of the Computer
  Software and Applications Conference (COMPSAC)}, vol.~1.\hskip 1em plus 0.5em
  minus 0.4em\relax IEEE, 2016, pp. 2--11.

\bibitem{intpti}
X.~Cheng, M.~Zhou, X.~Song, M.~Gu, and J.~Sun, ``{IntPTI: Automatic Integer
  Error Repair with Proper-Type Inference},'' in \emph{Proceedings of the
  International Conference on Automated Software Engineering (ASE)}.\hskip 1em
  plus 0.5em minus 0.4em\relax IEEE/ACM, 2017.

\bibitem{archer}
R.~Chinchani, A.~Iyer, B.~Jayaraman, and S.~Upadhyaya, ``{ARCHERR: Runtime
  Environment Driven Program Safety},'' in \emph{Proceedings of the European
  Symposium on Research in Computer Security (ESORICS)}, 2004.

\bibitem{cbmc}
E.~Clarke, D.~Kroening, and F.~Lerda, ``{A Tool for Checking {ANSI-C}
  Programs},'' in \emph{Proceedings of the International Conference on Tools
  and Algorithms for the Construction and Analysis of Systems (TACAS)}, 2004.

\bibitem{coker:integer}
Z.~Coker and M.~Hafiz, ``{Program Transformations to Fix {C} Integers},'' in
  \emph{Proceedings of the International Conference on Software Engineering
  (ICSE)}.\hskip 1em plus 0.5em minus 0.4em\relax IEEE/ACM, 2013, pp. 792--801.

\bibitem{cwe190}
\BIBentryALTinterwordspacing
M.~Corporation. {Integer Overflow or Wraparound}. [Online]. Available:
  \url{https://cwe.mitre.org/data/definitions/190.html}
\BIBentrySTDinterwordspacing

\bibitem{z3}
L.~De~Moura and N.~Bj{\o}rner, ``{Z3: An Efficient {SMT} Solver},'' in
  \emph{Tools and Algorithms for the Construction and Analysis of Systems
  (TACAS)}.\hskip 1em plus 0.5em minus 0.4em\relax Springer, 2008, pp.
  337--340.

\bibitem{ioc:icse}
W.~Dietz, P.~Li, J.~Regehr, and V.~Adve, ``{Understanding Integer Overflow in
  C/C++},'' in \emph{Proceedings of the International Conference on Software
  Engineering (ICSE)}.\hskip 1em plus 0.5em minus 0.4em\relax IEEE/ACM, 2012.

\bibitem{ioc}
W.~Dietz, P.~Li, J.~Regehr, and V.~Adve, ``{Understanding Integer Overflow in
  C/C++},'' \emph{Transactions on Software Engineering and Methodology
  (TOSEM)}, 2015.

\bibitem{cdt}
\BIBentryALTinterwordspacing
Eclipse. {Eclipse CDT}. [Online]. Available: \url{https://eclipse.org/cdt/}
\BIBentrySTDinterwordspacing

\bibitem{jface}
\BIBentryALTinterwordspacing
Eclipse. {Eclipse JFace}. [Online]. Available:
  \url{https://wiki.eclipse.org/JFace}
\BIBentrySTDinterwordspacing

\bibitem{ltk}
\BIBentryALTinterwordspacing
Eclipse. {The Language Toolkit: {A}n {API} for Automated Refactorings in
  {E}clipse-based {IDE}s}. [Online]. Available:
  \url{https://eclipse.org/articles/Article-LTK/ltk.html}
\BIBentrySTDinterwordspacing

\bibitem{gamma:patterns}
E.~Gamma, J.~Vlissides, R.~Johnson, and R.~Helm, \emph{{Design Patterns:
  {E}lements of Reusable Object-Oriented Software}}.\hskip 1em plus 0.5em minus
  0.4em\relax Addison-Wesley, 1994.

\bibitem{gcc}
\BIBentryALTinterwordspacing
GCC. {The GNU Compiler Collection}. [Online]. Available:
  \url{https://gcc.gnu.org//}
\BIBentrySTDinterwordspacing

\bibitem{sage}
P.~Godefroid, M.~Levin, and D.~Molnar, ``{Automated Whitebox Fuzz Testing},''
  in \emph{Proceedings of the Network and Distributed System Security Symposium
  (NDSS)}, 2008.

\bibitem{senx}
Z.~Huang, D.~Lie, G.~Tan, and T.~Jaeger, ``{Using Safety Properties to Generate
  Vulnerability Patches},'' in \emph{Proceedings of the Symposium on Security
  and Privacy (S\&P)}, 2019.

\bibitem{ibing:date}
A.~Ibing, ``{Architecture Description Language based Retargetable Symbolic
  Execution},'' in \emph{Proceedings of the Design, Automation and Test in
  Europe Conference and Exhibition (DATE)}, 2015.

\bibitem{ibing:hase}
A.~Ibing and A.~Mai, ``{A Fixed-Point Algorithm for Automated Static Detection
  of Infinite Loops},'' in \emph{Proceedings of the International Symposium on
  High Assurance Systems Engineering (HASE)}.\hskip 1em plus 0.5em minus
  0.4em\relax IEEE, 2015.

\bibitem{codan}
A.~Laskavaia, ``{Codan-C/C++ Static Analysis Framework for CDT},'' 2011.

\bibitem{sift}
F.~Long, S.~Sidiroglou-Douskos, D.~Kim, and M.~Rinard, ``{Sound Input Filter
  Generation for Integer Overflow Errors},'' in \emph{ACM Sigplan Notices},
  vol.~49, no.~1, 2014, pp. 439--452.

\bibitem{marinescu2012make}
P.~Marinescu and C.~Cadar, ``{{M}ake Test-zesti: {A} Symbolic Execution
  Solution for Improving Regression Testing},'' in \emph{Proceedings of the
  International Conference on Software Engineering (ICSE)}.\hskip 1em plus
  0.5em minus 0.4em\relax IEEE/ACM, 2012, pp. 716--726.

\bibitem{directfix}
S.~Mechtaev, J.~Yi, and A.~Roychoudhury, ``{DirectFix: {L}ooking for Simple
  Program Repairs},'' in \emph{Proceedings of the International Conference on
  Software Engineering (ICSE)}.\hskip 1em plus 0.5em minus 0.4em\relax
  IEEE/ACM, 2015.

\bibitem{angelix}
S.~Mechtaev, J.~Yi, and A.~Roychoudhury, ``{Angelix: Scalable Multiline Program
  Patch Synthesis via Symbolic Analysis},'' in \emph{Proceedings of the
  International Conference on Software Engineering (ICSE)}.\hskip 1em plus
  0.5em minus 0.4em\relax IEEE/ACM, 2016.

\bibitem{prefast}
\BIBentryALTinterwordspacing
Microsoft. {PREfast Analysis Tool}. [Online]. Available:
  \url{https://msdn.microsoft.com/en-us/library/ms933794.aspx}
\BIBentrySTDinterwordspacing

\bibitem{cwe192:coercion}
\BIBentryALTinterwordspacing
Mitre. {Integer Coercion Error}. [Online]. Available:
  \url{https://cwe.mitre.org/data/definitions/192.html}
\BIBentrySTDinterwordspacing

\bibitem{cwe191:underflow}
\BIBentryALTinterwordspacing
Mitre. {Integer Underflow (Wrap or Wraparound)}. [Online]. Available:
  \url{https://cwe.mitre.org/data/definitions/191.html}
\BIBentrySTDinterwordspacing

\bibitem{cwe680}
\BIBentryALTinterwordspacing
Mitre. {IO2BO: Integer Overflow to Buffer Overflow}. [Online]. Available:
  \url{https://cwe.mitre.org/data/slices/680.html}
\BIBentrySTDinterwordspacing

\bibitem{cwe197:truncation}
\BIBentryALTinterwordspacing
Mitre. {Numeric Truncation Error}. [Online]. Available:
  \url{https://cwe.mitre.org/data/definitions/197.html}
\BIBentrySTDinterwordspacing

\bibitem{cwe193:offbyone}
\BIBentryALTinterwordspacing
Mitre. {Off-by-one Error}. [Online]. Available:
  \url{http://cwe.mitre.org/data/definitions/193.html}
\BIBentrySTDinterwordspacing

\bibitem{cwe195:signed}
\BIBentryALTinterwordspacing
Mitre. {Signed to Unsigned Conversion Error}. [Online]. Available:
  \url{https://cwe.mitre.org/data/definitions/195.html}
\BIBentrySTDinterwordspacing

\bibitem{cwe194:sign}
\BIBentryALTinterwordspacing
Mitre. {Unexpected Sign Extension}. [Online]. Available:
  \url{https://cwe.mitre.org/data/definitions/194.html}
\BIBentrySTDinterwordspacing

\bibitem{cwe196:unsigned}
\BIBentryALTinterwordspacing
Mitre. {Unsigned to Signed Conversion Error}. [Online]. Available:
  \url{https://cwe.mitre.org/data/definitions/196.html}
\BIBentrySTDinterwordspacing

\bibitem{smartfuzz}
D.~Molnar, X.~C. Li, and D.~Wagner, ``{Dynamic Test Generation to Find Integer
  Bugs in x86 Binary {L}inux Programs},'' in \emph{Proceedings of the USENIX
  Security Symposium (USENIX Security)}, 2009, pp. 67--82.

\bibitem{Monperrus2015}
M.~Monperrus, ``{Automatic Software Repair: {A} Bibliography},'' \emph{{ACM
  Computing Surveys}}, vol.~51, no.~1, 2018.

\bibitem{prefix}
Y.~Moy, N.~Bjørner, and D.~Sielaff, ``{Modular Bug-finding for Integer
  Overflows in the Large: {S}ound, Efficient, Bit-precise Static Analysis},''
  \emph{MSR-TR-2009-57}, 2009.

\bibitem{tauCFI}
P.~Muntean, M.~Fischer, G.~Tan, Z.~Lin, J.~Grossklags, and C.~Eckert,
  ``{tauCFI: Type-Assisted Control Flow Integrity for x86-64 Binaries},'' in
  \emph{Proceedings of the Symposium on Research in Attacks, Intrusions, and
  Defenses (RAID)}, 2018.

\bibitem{safecomp}
P.~Muntean, V.~Kommanapali, A.~Ibing, and C.~Eckert, ``{Automated Generation of
  Buffer Overflow Quick Fixes Using Symbolic Execution and {SMT}},'' in
  \emph{Proceedings of the International Conference on Computer Safety,
  Reliability, and Security (SAFECOMP)}.\hskip 1em plus 0.5em minus 0.4em\relax
  LNCS, 2015.

\bibitem{issa}
P.~Muntean, M.~Rahman, A.~Ibing, and C.~Eckert, ``{{SMT}-constrained Symbolic
  Execution Engine for Integer Overflow Detection in {C} Code},'' in
  \emph{Proceedings of the Information Security for South Africa (ISSA)}.\hskip
  1em plus 0.5em minus 0.4em\relax IEEE, 2015.

\bibitem{castsan}
P.~Muntean, S.~Wuerl, J.~Grossklags, and C.~Eckert, ``{CastSan: Efficient
  Detection of Polymorphic C++ Object Type Confusions with LLVM},'' in
  \emph{Proceedings of the European Symposium on Research in Computer Security
  (ESORICS)}, 2018.

\bibitem{semfix}
H.~Nguyen, A.~Qi, A.~Roychoudhury, and S.~Chandra, ``{Semfix: {P}rogram Repair
  via Semantic Analysis},'' in \emph{Proceedings of the International
  Conference on Software Engineering (ICSE)}.\hskip 1em plus 0.5em minus
  0.4em\relax IEEE/ACM, 2013.

\bibitem{io:conditions}
F.~Pereira, R.~Rodrigues, and V.~Sperle~Campos, ``{A Fast and Low-overhead
  Technique to Secure Programs Against Integer Overflows},'' in
  \emph{Proceedings of the IEEE/ACM International Symposium on Code Generation
  and Optimization (CGO)}, 2013.

\bibitem{intflow}
M.~Pomonis, T.~Petsios, K.~Jee, M.~Polychronakis, and A.~D. Keromytis,
  ``{IntFlow: {I}mproving the Accuracy of Arithmetic Error Detection using
  Information flow Tracking},'' in \emph{Proceedings of the Annual Computer
  Security Applications Conference (ACSAC)}.\hskip 1em plus 0.5em minus
  0.4em\relax ACM, 2014, pp. 416--425.

\bibitem{tap}
S.~Sidiroglou-Douskos, E.~Lahtinen, and M.~Rinard, ``{Automatic Discovery and
  Patching of Buffer and Integer Overflow Errors},'' Massachusetts Institute of
  Technology, Tech. Rep. MIT-CSAIL-TR-2015-018, 2015.

\bibitem{diode}
S.~Sidiroglou-Douskos, E.~Lahtinen, N.~Rittenhouse, P.~Piselli, F.~Long,
  D.~Kim, and M.~Rinard, ``{Targeted Automatic Integer Overflow Discovery Using
  Goal-directed Conditional Branch Enforcement},'' in \emph{ACM Sigplan
  Notices}, vol.~50, no.~4, 2015, pp. 473--486.

\bibitem{codephage}
S.~Sidiroglou-Douskos, E.~Lahtinen, F.~Long, P.~Piselli, and M.~Rinard,
  ``{Automatic Error Elimination by Multi-application Code Transfer},''
  MIT-CSAIL-TR-2014-024, Tech. Rep., 2014.

\bibitem{inteq}
H.~Sun, X.~Zhang, Y.~Zheng, and Q.~Zeng, ``{IntEQ: Recognizing Benign Integer
  Overflows via Equivalence Checking Across Multiple Precisions},'' in
  \emph{Proceedings of the International Conference on Software Engineering
  (ICSE)}.\hskip 1em plus 0.5em minus 0.4em\relax IEEE/ACM, 2016, pp.
  1051--1062.

\bibitem{osiris}
C.~F. Torres, J.~Schuette, and R.~State, ``{Osiris: Hunting for Integer Bugs in
  {E}thereum Smart Contracts},'' in \emph{Proceedings of the Annual Computer
  Security Applications Conference (ACSAC)}.\hskip 1em plus 0.5em minus
  0.4em\relax ACM, 2018, pp. 416--425.

\bibitem{juliet}
\BIBentryALTinterwordspacing
{U.S. National Institute of Standards and Technology (NIST)}. {Juliet {T}est
  {S}uite v1.2 for {C}/{C}++}. [Online]. Available:
  \url{https://samate.nist.gov/SRD/testsuite.php}
\BIBentrySTDinterwordspacing

\bibitem{soupint}
T.~Wang, C.~Song, and W.~Lee, ``{Diagnosis and Emergency Patch Generation for
  Integer Overflow Exploits},'' in \emph{International Conference on Detection
  of Intrusions and Malware, and Vulnerability Assessment (DIMVA)}.\hskip 1em
  plus 0.5em minus 0.4em\relax Springer, 2014, pp. 255--275.

\bibitem{intscope}
T.~Wang, T.~Wei, Z.~Lin, and W.~Zou, ``{Int{S}cope: Automatically Detecting
  Integer Overflow Vulnerability in X86 Binary Using Symbolic Execution},'' in
  \emph{Network and Distributed System Security Symposium (NDSS)}, 2009.

\bibitem{uqbtng}
R.~Wojtczuk, ``{UQBTng: A Tool Capable of Automatically Finding Integer
  Overflows in {W}in32 Binaries},'' \emph{Chaos Communication Congress (22C3)},
  2005.

\bibitem{search:repair}
K.~Yalin, K.~T. Stolee, C.~Le~Goues, and Y.~Brun, ``{Repairing Programs with
  Semantic Code Search},'' in \emph{Proceedings of the International Conference
  on Automated Software Engineering (ASE)}.\hskip 1em plus 0.5em minus
  0.4em\relax IEEE/ACM, 2015.

\bibitem{toolx}
B.~Zhang, C.~Feng, B.~Wu, and C.Tang, ``{Detecting Integer Overflow in
  {W}indows Binary Executables based on Symbolic Execution},'' in
  \emph{Proceedings of the International Conference on Software Engineering,
  Artificial Intelligence, Networking and Parallel/Distributed Computing
  (SNPD)}, 2016.

\bibitem{intpatch}
C.~Zhang, T.~Wang, T.~Wei, Y.~Chen, and W.~Zou, ``{IntPatch: Automatically Fix
  Integer-Overflow-to-Buffer-Overflow Vulnerability at Compile-Time},'' in
  \emph{Proceedings of the European Symposium on Research in Computer Security
  (ESORICS)}, 2010, pp. 71--86.

\bibitem{indio}
Y.~Zhang, X.~Sun, Y.~Deng, L.~Cheng, S.~Zeng, Y.~Fu, and D.~Feng, ``{Improving
  Accuracy of Static Integer Overflow Detection in Binary},'' in
  \emph{Proceedings of the International Symposium on Research in Attacks,
  Intrusions, and Defenses (RAID)}, 2015.

\end{thebibliography}

\newpage
\begin{IEEEbiography}[{\includegraphics[width=1in,height=1.25in,clip,keepaspectratio]{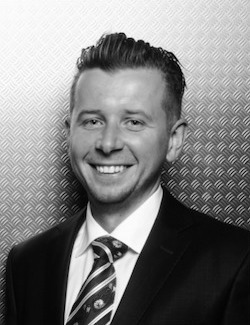}}]{Paul Muntean}
is Senior Cyber Security Engineer at Novartis Institute for Biological Research (NIBR) in Basel, Switzerland.
In the past, he has been visiting researcher at EPFL, Lausanne, Switzerland; and The Ohio State University, USA. He was a Ph.D. student at the Technical University of Munich, Germany. His research focuses on
the development of techniques and tools for
program analysis in order to detect security vulnerabilities and generate program repairs automatically.
\end{IEEEbiography}

\begin{IEEEbiography}[{\includegraphics[width=1in,height=1.25in,clip,keepaspectratio]{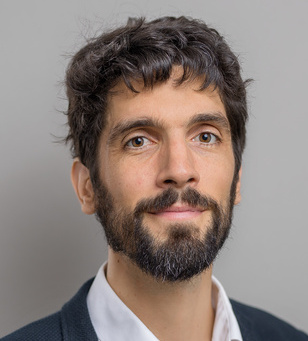}}]{Martin Monperrus}
is Professor of Software Technology at KTH Royal Institute of Technology, Sweden. He was previously associate professor at the University of Lille and adjunct researcher at Inria, France. He received a Ph.D. from the University of Rennes, France; and a Master's degree from the Compiègne University of Technology, France. His research lies in the field of software engineering with a current focus on automatic program repair, program hardening and chaos engineering. 
\end{IEEEbiography}

\begin{IEEEbiography}[{\includegraphics[width=1in,height=1.25in,clip,keepaspectratio]{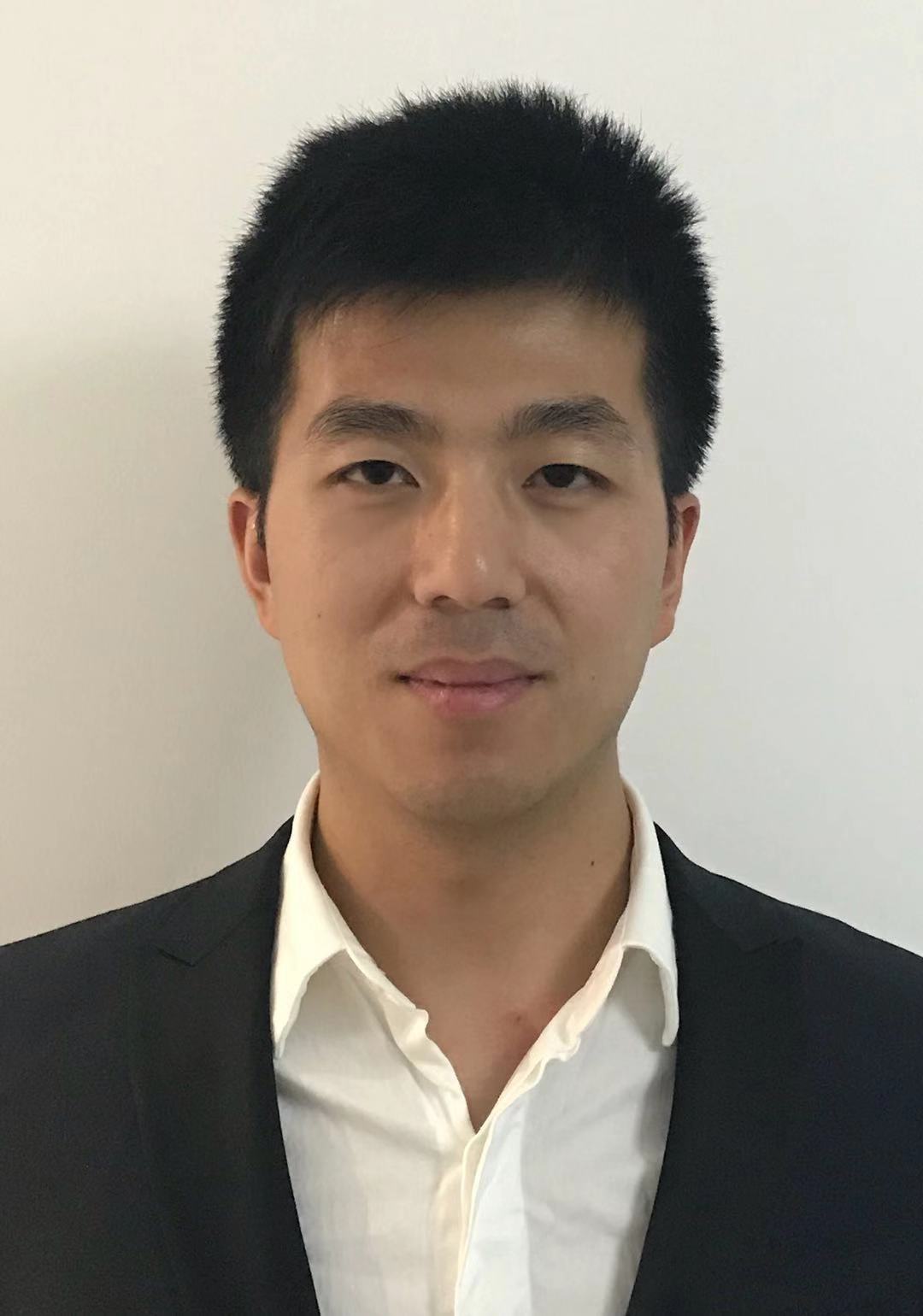}}]{Hao Sun}
received his B.Sc. and Ph.D. degrees in computer science from Nanjing University, China. He was a visiting scholar at Purdue Uninversity, USA; and worked as software engineer at Alibaba Group in Hangzhou, China. His main research interests include program analysis, software security and software testing. He has successfully published research results in venues such as ICSE, ASE, and AsiaCCS.
\end{IEEEbiography}

\begin{IEEEbiography}[{\includegraphics[width=1in,height=1.25in,clip,keepaspectratio]{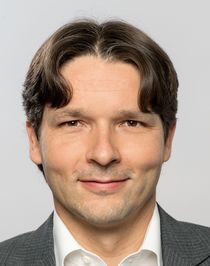}}]{Jens Grossklags} is an Associate Professor at the Technical University of Munich. Previously, he served as a post-doctoral research associate at Princeton University, and as Haile Family Early Career Professor at the Pennsylvania State University. He received his Ph.D. from the School of Information, UC Berkeley. In his interdisciplinary research agenda, he is investigating security and privacy challenges from the theoretical and practical perspectives. He is senior member of the ACM and IEEE.
\end{IEEEbiography}

\begin{IEEEbiography}[{\includegraphics[width=1in,height=1.25in,clip,keepaspectratio]{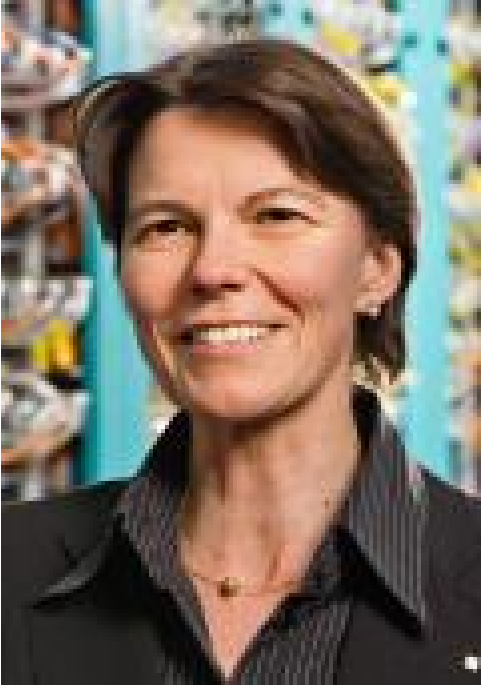}}]{Claudia Eckert}
is Full Professor at the Technical University of Munich, Germany. 
She is the director of the Fraunhofer Institute for Applied and Integrated Security (AISEC). She advises government
departments and the public sector at national and international levels in the development of research strategies and the implementation of security concepts.
Her research and teaching activities are centered around the topics of information security, operating systems, middleware, and communication networks.

\end{IEEEbiography}

\end{document}